# The land use-climate change-biodiversity nexus in European islands stakeholders

Aristides Moustakas[1,*], Irene Christoforidi[2], George Zittis[3], Nazli Demirel[4], Mauro Fois[5], Savvas Zotos[6], Eirini Gallou[7], Valentini Stamatiadou[8], Elli Tzirkalli[6], Christos Zoumides[9], Kristina Košić[10], Aikaterini Christopoulou[11], Aleksandra Dragin[10], Damian Łowicki[12], Artur Gil[13], Bruna Almeida[14], Panos Chrysos[15], Mario V. Balzan[16], Mark D.C. Mansoldo[16], Rannveig Ólafsdóttir[17], Cigdem Kaptan Ayhan[18], Lutfi Atay[19], Mirela Tase[20], Vladimir Stojanović[10], Maja Mijatov Ladičorbić[10], Juan Pedro Díaz[21], Francisco Javier Expósito[21], Sonia Quiroga[22], Miguel Ángel Casquet Cano[22], Haoran Wang[22], Cristina Suárez[23], Paraskevi Manolaki[6], Ioannis N. Vogiatzakis[6]

1. Natural History Museum of Crete, University of Crete, Heraklion, Crete, Greece

2. Department of Agriculture, Hellenic Mediterranean University, 71410 Heraklion, Greece

3. Climate and Atmosphere Research Center (CARE-C), The Cyprus Institute, Nicosia, Cyprus

4. Institute of Marine Sciences and Management, Istanbul University, Istanbul, Türkiye

5. Department of Life and Environmental Sciences, University of Cagliari, Cagliari, Italy

6. Faculty of Pure and Applied Sciences, Open University of Cyprus, Latsia, Cyprus

7. Centre for Sustainable Development, University of Strathclyde, Glasgow, UK.

8. Department of Marine Sciences, University of the Aegean, Mytilene, Greece

9. Energy, Environment & Water Research Center (EEWRC), The Cyprus Institute, Nicosia, Cyprus

10. Department of Geography, Tourism and Hotel Management, University of Novi Sad, Novi Sad, Serbia

11. Department of Economics and Sustainable Development, Harokopio University, Athens, Greece

12. Faculty of Human Geography and Planning, Adam Mickiewicz University, Poznań, Poland

13. IVAR - Research Institute for Volcanology and Risk Assessment, University of the Azores, Azores, Portugal

14. NOVA Information Management School (NOVA IMS), Universidade Nova de Lisboa, Lisbon, Portugal

15. Landscape Architecture Development, Santorini, Greece

16. Ecostack Innovations Limited, Kordin Business Incubation Centre, Kordin, Malta

17. Department of Geography and Tourism Studies, Institute of Life and Environmental Sciences, University of Iceland, Iceland.

18. Department of Landscape Architecture, Canakkale Onsekiz Mart University, Canakkale, Türkiye

19. Department of Travel Management and Guidence, Canakkale Onsekiz Mart University, Canakkale, Türkiye

20. Department of Tourism, Aleksander Mosiu University, Durres, Albania

21. Grupo de Observación de la Tierra y la Atmósfera (GOTA), Departamento de Física, Universidad de La Laguna (ULL), San Cristóbal de La Laguna, Canary Islands, Spain

22. Department of Economic Analysis and Quantitative Economics, Complutense University of Madrid, Madrid, Spain

23. Department of Economics. Universidad de Alcalá. Alcalá de Henares, Madrid, Spain

*Corresponding author: arismoustakas@uoc.gr; arismoustakas@gmail.com

**Abstract**

To promote climate adaptation and mitigation strategies, it is crucial to understand the perspectives and knowledge gaps of stakeholders involved in functions affected by or addressing land use and climate changes. A large number of stakeholders across 21 European islands were consulted regarding their views on climate change and land use change issues affecting ecosystem services on their island. Climate change characteristics perceptions included variables such as temperature, precipitation, humidity, extremes, and wind. Land use change characteristics perceptions included deforestation, coastal degradation, habitat protection, renewable energy facilities, wetlands and other variables. Other environmental and societal problem perceptions such as invasive species, water or energy scarcity, problems in infrastructures or austerity were also included. Climate and land use change impact perceptions were analysed with machine learning to quantify their importance on the perception outcome. For example if a stakeholder perceives that pollution, coastal degradation, deforestation, precipitation decrease, and increase of humidity are occurring on the island, and austerity is the biggest problem how likely is that the impact of climate change or land use change will be quantified by the stakeholder as negative, unclear, neutral, or positive? The predominant climatic change characteristic is related with temperature, and the predominant land use change characteristic with deforestation. Water-related problems are top priorities for stakeholders. Energy-related problems, such as energy deficiency but also wind and solar energy facilities problems, rank high as combined climate change and land use change risks. Stakeholders generally perceive climate change impacts on ecosystem services as negative, with natural habitat destruction and biodiversity loss identified as the top variables. Land use change impacts are also negative but also more complex to explain, with a higher number of explanatory variables associated with the impact outcome. Stakeholders have common perceptions regarding climate change and land use change impacts on the benefits of biodiversity despite the geographic disparity. Stakeholders differentiate between factors related to climate change impacts and land use change impacts. Water, energy, and renewable energy related issues pose serious concerns to island stakeholders and management measures are needed to address them.

**Keywords**

Impact assessment; climate change; land use change; islands; ecosystem services; machine learning

## Introduction

The rapid pace of global change, characterized mainly by climatic and land-use changes, presents significant challenges to biodiversity conservation and sustainable development (Pörtner et al., 2023; WWF 2020). Islands make up only a small portion of the Earth's surface, yet they are home to over one-third of the global biodiversity (Pichot et al., 2024; Steibl et al., 2024). Ecosystem services, including provisioning, regulating, mediating, and cultural functions, are underpinned by essential ecosystem processes like soil formation, photosynthesis, pollination, and nutrient cycling (Elmqvist et al., 2012). These services are also heavily dependent on climate conditions, land use, and anthropogenic disturbances (Zittis *et al.*, 2025). However, significant knowledge gaps remain regarding island ecosystem services and their response to climate and land-use changes, hindering both basic understanding as well as science to policy assessments (Mycoo and Roopnarine 2024; Solé Figueras et al., 2024). European islands face unique challenges regarding ecosystem services, including vulnerability to climate change, habitat loss, pollution, and invasive species, all of which can impact the flow of vital services like clean water, food production, and recreation (Vogiatzakis *et al.*, 2023). Challenges are often exacerbated by factors like tourism development, land-use changes, and resource extraction (Aretano *et al.*, 2013). These facts and that island ecosystems are particularly vulnerable makes them valuable case studies for exploring the interplay within the land use–climate change–biodiversity nexus (Moustakas *et al.*, 2025).

Key climatic parameters affecting ecosystems include temperature, precipitation, humidity, wind, extremes, hydrological cycle components, and oceanic properties. Climate change can also impact mental health, wellbeing, and sense of safety (Clayton and Swim 2025) due, for example, to extreme weather events that damage both natural habitats and infrastructure (Handmer et al., 2012). These habitats and infrastructure may be critical for the island social, economic, and environmental functioning as there may be no other of the kind (Martin del Campo et al., 2023; McEvoy et al., 2024). Considering that many islands are attractive tourist destinations, climate change may be a critical factor in modulating the length and peaks of tourist season (Becken and Wilson 2013), for example, through higher temperatures in spring and autumn or more heatwaves in summer (Hernandez et al., 2018). In addition, tourism increases the total demand for water (Becken 2014), making it difficult to meet the needs (Falkland 1999) already in deficit from the effects of climate change and increasing demand from farmers and primary productivity (Kourgialas 2021).

Land use changes, including urban expansion, mining, deforestation, coastal zone degradation, and wetland modification, frequently result in habitat loss, fragmentation, and biodiversity decline (Haines-Young 2009). These changes can undermine climate change mitigation and adaptation efforts, often interacting synergistically with climatic impacts, though such interactions are poorly understood (De Chazal and Rounsevell, 2009; He et al., 2019). While certain land-use changes, such as habitat protection, rewilding, and reforestation, can enhance biodiversity and ecosystem services, land-use change remains the primary threat to biodiversity on a global scale (IPBES, 2019; WWF, 2020, 2022). Notably, these conclusions are predominantly based on terrestrial data and fail to capture the unique dynamics of island ecosystems, where land-use pressures often concentrate in coastal zones due to high demand for housing and tourism infrastructure (Vogiatzakis et al., 2023). Seasonal energy demands further exacerbate these pressures, leading to the installation of wind and solar energy facilities (Mauger *et al.*, 2024). Although these installations promote renewable energy, they also contribute to habitat fragmentation and biodiversity loss by occupying agricultural and natural land areas (Kuang et al., 2016). The European Commission's "renewable islands for 2030" initiative has confirmed 30 islands and groups aiming for energy independence by 2030, despite inadequate spatial planning and insufficient stakeholder consultation (Cuka, 2025).

Biodiversity and ecosystem services are influenced by various additional social and economic factors (Kopnina et al., 2024). These may include demographic growth, technological advancements intensifying or relaxing natural resource exploitation, and pollution. In addition, invasive alien species, diseases, and pests can have pronounced effects on island ecosystems (Bjarnason et al., 2017; Thaman 2002). Overtourism generates pressure in island infrastructure such as roads and airports, housing, cost of living and waste, acting as a deterring factor for living or working in such islands stimulating debates about sustainable practices and development (Fernández et al., 2024; Kelman 2022). Admittedly not everything can be attributed to climatic or land use changes alone. For instance, in one of the world's lowest-lying island nations (Marshall Islands) that are vulnerable to sea level rise and flooding, the majority of their population identified that education, healthcare, employment, and family visits outranked climate change and sea level rise as the main migration motive (van der Geest et al., 2020). Similar results have been recorded in other islands where financial austerity is common (Butler et al., 2014). Thus, climatic and land use factors need to be analysed in conjunction with social, financial, and other environmental factors in order to quantify their overall importance.

Perception is a subjective assessment of a concept or sensation, influenced by interests, personal growth, and the environment with profound variations between locations and individuals or groups within locations (Antronico et al., 2020; Dhar et al., 2023). In addition, the response to perception is influenced by cultural or monetary aspects and worldviews, such as the potential role of individual actions, anxiety over future climate scenarios, or the inevitability of climate change (Salas Reyes et al., 2021). Thus, an interdisciplinary approach including social, climatic, environmental, and biological sciences is needed (Balzter et al., 2023). In addition, the scientific methods deployed need to account for the complexity of the problem investigating simultaneously several potential partly correlated variables and identifying their relevant contribution to the perception outcome (Fisher et al., 2019). Questionnaires are a popular tool in social surveys to address information about stakeholder perceptions, knowledge, and beliefs that are widely used for perception analyses, providing a heuristic function that allows information to be sorted and retrieved (Otto-Banaszak et al., 2011; Tourlioti et al., 2024).

The aim of the present study was to collect information on climate change and land use change related challenges issues from a local stakeholder perspective, and identify the main factors affecting European islands (Vogiatzakis et al., 2023). Several studies on nexus approaches explore interconnections across sectors (Authier et al., 2024; Dale et al., 2011; Newman et al., 2020; Nie et al., 2019), yet this research stands out by addressing the interplay of climate change, land use, and biodiversity benefits specifically within the climatically, biodiversity, culturally, economically, and geographically diverse context of European islands. Our research addresses the land use-climate change-biodiversity nexus by analyzing perceptions of stakeholders residing on islands. None have focused on such a broad and diverse spectrum of islands and archipelagos. This work deploys data from questionnaires distributed in over 21 European islands or groups of islands and across a large number of stakeholders in order to assess the perception of problems and associated factors regarding the land use changes-climatic changes-biodiversity nexus (Dale et al., 2011; Rasmus et al., 2024; Ruiz et al., 2023). Data were analysed with machine learning in order to simultaneously handle together a large number of variables and quantify their contribution to the perception of environmental problems in island societies (Fisher et al., 2019; Ruiz et al., 2023).

Stakeholders included are permanent inhabitants of the island studied or individuals that their primary financial activity is associated with presence on the island, from diverse backgrounds. The research questions addressed are: (1) What are the climatic variables that best quantify climate change? (2) What are the land use or land cover characteristics that best quantify land use change? (3) What are the climate- and land use change-induced problems on ecosystem services and how severe are they? (4) In addition to climatic changes and land use changes, what are the main additional environmental, societal, and economic problems and how severe are they?  (5) What are the impacts of climate change on the benefits of biodiversity? Can impact perceptions outcomes be explained by climate and land use changing variables, climate change and land use change induced problems, together with the additional environmental, economical, and societal problems occurring on the island? How important is each variable on the impact outcome perception? (6) What are the impacts of land use change on the benefits of biodiversity? Can impact perceptions outcomes be explained by climate and land use changing variables, climate change and land use change induced problems, together with the additional environmental, economical, and societal problems occurring on the island? How important is each variable on the impact outcome perception?

**Methods**

The study included 737 stakeholders from 21 different islands or archipelagos across 12 European countries (Fig. 1). Stakeholder perceptions were recorded between February to April 2024. Stakeholders included are either permanent inhabitants of the island or individuals that their main financial activity is related with continues presence on the island. Islands were selected opportunistically based on the willingness of an academic researcher to participate in the study and collect stakeholder data on a European island that is within their expertise and research interests. Each academic researcher participating acted as mediator between the island stakeholders, was present on the study island and responsible for identifying stakeholders and contacting them.

Regarding stakeholders within each island, diverse professional groups were opportunistically contacted per island, so as to cover a wide range of professional activities including eight broad groups: (1) public authority/policy makers, (2) professionals in primary sector (agriculture, fisheries, stock raising, forestry), (3) technicians and associated professionals, (4) professionals in the food industry, (5) professionals in the financial sector, (6) academics/researchers, (7) professionals in tourism, or (8) other, including among others house working individuals, students residing on the island, or unemployed. Sampling across professions incorporates heterogeneity in the perceptions and ecosystem services of interest. Call for participation was conducted via the online platforms and dissemination tools of the research program that funds this research. This aimed mainly at academic stakeholders residing on the island (other than the academic collecting the data for the island). In addition the questionnaires were emailed on mailing lists, posted on social media web pages, and official email addresses of various professional island groups, and authorities. This was achieved by sending them an excel file or a link to an online Google form with the questions and feasible scores (Table 1). In addition, face-to-face questionnaire completion with stakeholders was conducted. Online answers accounted for 42% of stakeholders, while completed in person accounted for 58% of stakeholders.

Prior to filling in a questionnaire from each stakeholder, a basic definition of "Ecosystem Services" was provided to be on a safe side that the concept is understood. Ecosystem services were defined as “the benefits that humans derive from biodiversity and ecosystems, including both the goods and services that nature provides. These services, which are essential for human well-being and economic activity, range from the production of food and clean water to the regulation of climate and disease”. Some practical examples followed. Stakeholders were asked to indicate their perceptions on a binary and some on a five-point ‘Likert-type’ scale (Feliciano *et al.*, 2017) allowing for ranking observed effects of climate change for example (with 0 counting for Don’t know). The application of ranking for assessing level of perceived significance and the relevant answer options provided are described in Table 1.

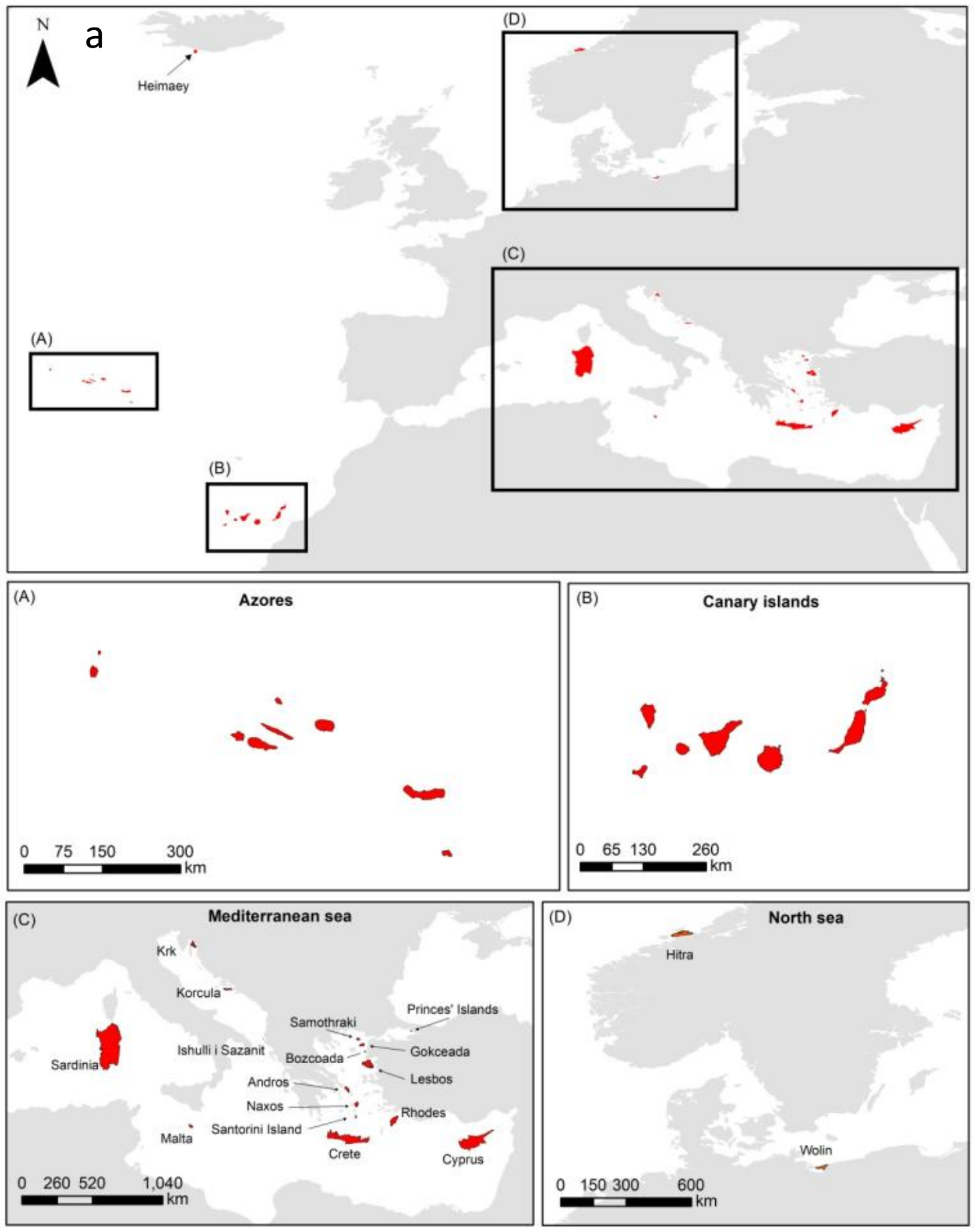

b

| Island | Country | Size (km$^2$) | Stake holders |
|---|---|---|---|
| Sazan | Albania | 5.70 | 30 |
| Krk | Croatia | 405.80 | 30 |
| Korcula | Croatia | 279.00 | 30 |
| Cyprus | Cyprus | 9,250.00 | 60 |
| Crete | Greece | 8,336.00 | 30 |
| Andros | Greece | 379.20 | 30 |
| Santorini | Greece | 76.19 | 30 |
| Rhodes | Greece | 1,400.68 | 30 |
| Naxos | Greece | 429.79 | 69 |
| Samothraki | Greece | 180.36 | 10 |
| Lesvos | Greece | 1,633.00 | 30 |
| Heimaey | Iceland | 13.40 | 10 |
| Sardinia | Italy | 24,090.00 | 91 |
| Malta | Malta | 316.00 | 34 |
| Hitra | Norway | 571.50 | 30 |
| Wolin | Poland | 265.00 | 31 |
| Azores | Portugal | 759.00 | 31 |
| Canary Islands | Spain | 7,493.00 | 41 |
| Princes' Island | Turkey | 10.60 | 30 |
| Gokceada | Turkey | 286.00 | 30 |
| Bozcaada | Turkey | 37.60 | 30 |

***Figure 1. a.*** *European islands location. Each island included is depicted in red colour.* ***b.*** *Country of the island, surface area, and number of stakeholders.*

*Questionnaire design*

The questionnaire consisted of pre-defined, close-ended questions designed to quantify island stakeholders' perceptions of the land use change–climate change–biodiversity benefits nexus (Otto-Banaszak et al., 2011; Tourlioti et al., 2024). Responses were either binary (YES/NO) or based on ranking problems on a 0–4 scale (0 = N/A, 4 = highest intensity). The survey was conducted in twelve languages (English, Spanish, Italian, Greek, Albanian, Turkish, Croatian, Portuguese, Norwegian, Polish, Maltese, and Icelandic). Participation was voluntary and anonymous, and all respondents were adults (>18 years). A minimum of 30 participants per island was targeted; however, only 10 were recruited in Heimaey and Samothraki (Fig. 1). Even in those islands a heterogenous sample regarding professions of stakeholders was achieved.

*Climate change characteristics*

Binary questions (0 = No, 1 = Yes) captured perceptions of changes in: (1) temperature (increase, decrease, extreme events, heat island effects), (2) precipitation (increase, decrease, variability/extremes), (3) wind (increase/storms), (4) ocean acidification, (5) sea level rise, and (6) humidity (Table 1). Multiple responses were permitted with examples provided in the questionnaire.

*Land use change characteristics*

Binary questions (0 = No, 1 = Yes) assessed perceived changes in: (1) deforestation (e.g., fire, tourist infrastructure, agriculture), (2) reforestation, (3) rewilding, (4) wetland modification, (5) coastal zone degradation, (6) habitat protection, (7) wind/solar energy facilities, (8) mining, and (9) urban expansion (Table 1). Multiple responses were permitted, with examples provided in the questionnaire.

*Climate and land use change problems*

Stakeholders ranked 14 potential problems (0–4 scale; 0 = N/A, 1: Not so much, 2: Relatively yes 3: Yes, observable change to 4: Yes, very much climate -induced change), including changes in tourist destinations, damage to landmarks, food risks, ecosystem destruction, urban changes, new diseases, invasive species, carbon sequestration, pollination, overconsumption, soil erosion, water issues, ecosystem recreation, and infrastructure damage (Table 1). Examples were provided to ensure participants understand the question in a similar way and can interpret it in similar way when answering (eg. For water issues, we mentioned water security, salinity levels, pollution etc and for 'ecosystem recreation' we provided explanation in the form of benefits from visiting nature and enjoying activities outdoors).

*Overall problems*

A second ranking question (0–4 scale; 0 for Don't Know, 1 for having a small effect, 2 having a moderate effect, 3 having an important, to 4 having a very big/significant effect) evaluated overall perceived problems, including separate scores for climate change, land use change, population growth, economic growth, pollution, resource extraction and degradation, policy changes, biodiversity overexploitation, and financial problems (Table 1). This allowed us to capture residents' participants observed effects while assigning them a relative level of presence on each island context.

*Impacts on ecosystem services*

Stakeholders evaluated the impact of (a) climate change and (b) land use change on biodiversity-related ecosystem services using four options: negative (1), unclear (2), neutral (3), or positive (4). To ensure participants understand the meaning of ecosystem services, we provided a common explanation to the term services, as the benefits stemming from ecosystems and their use for supporting human and planetary wellbeing. Responses were mutually exclusive and treated as ordinal values, with lower scores reflecting stronger negative impacts.

***Table 1**. Variables recorded and their feasible answers or scores*

| Category | Var Nr | Var Name | Value | | | | |
|---|---|---|---|---|---|---|---|
| **(i) Climate change variables** | | | | | | | |
| **Question (i) What are the climate characteristics changing in your island?** | 1 | Temperature | Yes/No | | | | |
| | 2 | Precipitation | Yes/No | | | | |
| | 3 | Wind increase/storms | Yes/No | | | | |
| | 4 | Ocean acidification | Yes/No | | | | |
| | 5 | Sea level rise | Yes/No | | | | |
| | 6 | Humidity increase | Yes/No | | | | |
| **(ii) Land use / land cover change variables** | | | | | | | |
| **Question (ii) What are the observable land use / land cover changes in your island?** | 7 | Deforestation | Yes/No | | | | |
| | 8 | Reforestation | Yes/No | | | | |
| | 9 | Rewilding | Yes/No | | | | |
| | 10 | Wetland modification | Yes/No | | | | |
| | 11 | Coastal degradation | Yes/No | | | | |
| | 12 | Habitat protection | Yes/No | | | | |
| | 13 | Renewable energy facility issues | Yes/No | | | | |
| | 14 | Mining related issues | Yes/No | | | | |
| | 15 | Urban expansion related issues | Yes/No | | | | |
| **(iii)Ranking overall problems** | Values ranking the observed effects of problems on the island context, from 0 for Don't Know, 1 for having a small effect, 2 having a moderate effect, 3 having an important, to 4 having a very big/significant effect | | | | | | |
| **(iii) Rank the effects of these problems on your island** | 16 | Climate change | 0 | 1 | 2 | 3 | 4 |
| | 17 | Land use change | 0 | 1 | 2 | 3 | 4 |
| | 18 | Population growth | 0 | 1 | 2 | 3 | 4 |
| | 19 | Economic growth | 0 | 1 | 2 | 3 | 4 |
| | 20 | Pollution | 0 | 1 | 2 | 3 | 4 |
| | 21 | Resource extraction | 0 | 1 | 2 | 3 | 4 |
| | 22 | Policy changes | 0 | 1 | 2 | 3 | 4 |
| | 23 | Nature overexploitation | 0 | 1 | 2 | 3 | 4 |
| | 24 | Austerity | 0 | 1 | 2 | 3 | 4 |
| **(iv) Climate change & land use change problems.** | Values ranking the perceived presence of each problem induced by climate change, from 0 for Don't Know, 1: Not so much, 2: Relative change yes 3: Yes, observable change to 4: Yes, very much climate -induced change. | | | | | | |
| **Question (iv) How would you assess the main climate and land use change induced problems in your island? (0-4)** | 25 | Tourism | 0 | 1 | 2 | 3 | 4 |
| | 26 | Damages in habitats, protected areas or monuments etc | 0 | 1 | 2 | 3 | 4 |
| | 27 | Primary sector (Agriculture, fisheries etc) | 0 | 1 | 2 | 3 | 4 |
| | 28 | Ecosystem Destruction/Biodiversity loss | 0 | 1 | 2 | 3 | 4 |
| | 29 | Changes in urban environment | 0 | 1 | 2 | 3 | 4 |
| | 30 | New diseases | 0 | 1 | 2 | 3 | 4 |
| | 31 | Invasive species | 0 | 1 | 2 | 3 | 4 |
| | 32 | Carbon sequestration | 0 | 1 | 2 | 3 | 4 |
| | 33 | Pollination | 0 | 1 | 2 | 3 | 4 |
| | 34 | Energy related issues | 0 | 1 | 2 | 3 | 4 |
| | 35 | Soil erosion | 0 | 1 | 2 | 3 | 4 |
| | 36 | Water related issues | 0 | 1 | 2 | 3 | 4 |
| | 37 | Ecosystem recreation | 0 | 1 | 2 | 3 | 4 |
| | 38 | Infrastructure damages | 0 | 1 | 2 | 3 | 4 |
| **Climate change impact on the benefits of biodiversity.** | 39 | Climate change impact on ecosystem services | Negative | Unclear | Neutral | Positive | |
| **Land use change impact on the benefits of biodiversity** | 40 | Land use change impact on ecosystem services | Negative | Unclear | Neutral | Positive | |

*Impact analysis on ecosystem services*

We sought to explain what drives perceptions (negative, unclear, neutral, or positive) regarding the impact of climate change on the benefits of biodiversity. To do so we combined all data regarding climate characteristics changing, land use characteristics changing, problems related with climate change and land use change, and overall societal, economic, and other environmental problems occurring as perceived by the stakeholders. Perceptions of island stakeholders analysed here derive from diverse stakeholder backgrounds, islands, countries, isolation factors, climate & land use characteristics and other societal, economic, and environmental problems. The data involves a large number of variables and perceptions potentially correlated across stakeholders' background, island geography, or country. For the analysis of impact perceptions, we used Random Forest (RF) classifiers, a machine learning technique (Breiman 2001). RFs are among the most efficient analytic tools for extracting information in noisy, complex, potentially correlated, and high-dimensional datasets such as the one deployed here and have been applied to a wide range of environmental topics (Daliakopoulos et al., 2017; Fisher et al., 2024; Moustakas and Davlias 2021).

In order to quantify stakeholder perceptions regarding climate change impacts on the benefits of biodiversity, the dependent variable included climate change impact on ecosystem services (variable 39, feasible scores 'negative = 1', 'unknown = 2', 'neutral = 3', or 'positive = 4'.). RFs were fit with independent explanatory variables that included (i) the climate change characteristics (variables 1-6), (ii) land use / land cover change characteristics (variables (7-15), (iii) climate change and land use change problems (variables 18-24; variables 16-17 not used), and (iv) overall problems (variables 25-38) as independent variables.

The analysis was replicated with the same independent explanatory variables as above (variables 1-38 excluding variables 16-17), and as dependent variable, land use change impacts on ecosystem services (variable 40, feasible scores also ‘negative = 1’, ‘unknown = 2’, ‘neutral = 3’, or ‘positive = 4’).

We trained and tested RF classifiers, using 10 different random states. All binary variables (variables 1-15) were treated as categorical predictors, while variables 18-38 as continuous predictors. The data split partitioning approach explored included a random partition of 70% of stakeholders for training and 30% for testing model outputs (Alif and Fahrudin 2024). Sensitivity analysis performed by changing the training to testing data partition (Moreno-Alcayde et al., 2024) by up to 10% resulted in similar or almost identical results. We did not employ multiple machine learning methods simultaneously, ensemble learning, and compare their accuracy (Sakti et al., 2024) as the objective of this study is to explicate rather than predict in novel circumstances, or to evaluate the most effective predictive tools and their accuracy.

The accuracy is defined as the number of correct bin classifications divided by the total number of instances in the dataset. In order to quantify the relative importance of each factor in the impact assessment outcome a relative variable importance chart was deployed (Venkateswarlu and Anmala 2024). The relative variable importance chart depicts the predictors in descending order of their impact on model enhancement from all the basis functions for a predictor. The relative variable importance is used to standardize the importance values for easier interpretation (Venkateswarlu and Anmala 2024). Relative importance is defined as the percentage improvement relative to the most important predictor, which has an importance of 100%. The relative importance is calculated by dividing each variable's importance score by the largest value of the variables, then multiplying by 100%.

## Results

### *Climate change characteristics*

Most stakeholders identified temperature, followed by precipitation, wind, and humidity as the most predominant climate characteristics under change on their island (Fig. 2a). Acidification and sea level rise were reported least frequently in the European islands investigated (Fig. 2a).

### *Land use change characteristics*

Stakeholders ranked deforestation as the most common land-use change process, followed by coastal degradation and urban expansion (Fig. 2b). Renewable energy facilities (solar and wind) were the fourth most reported issue, with insufficient habitat protection ranked fifth (Fig. 2b). Rewilding or reforestation was also considered a concern (Fig. 2b).

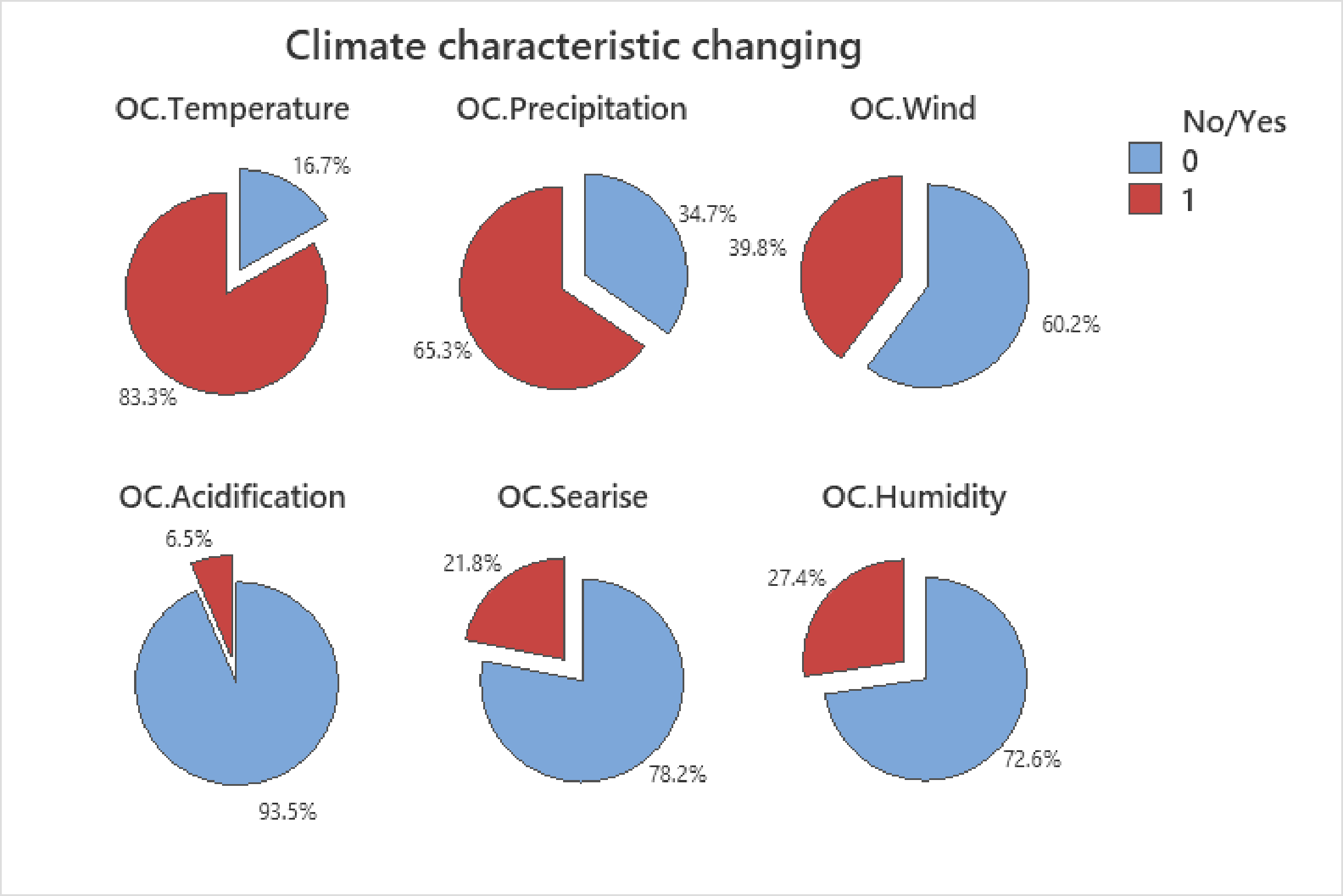

a

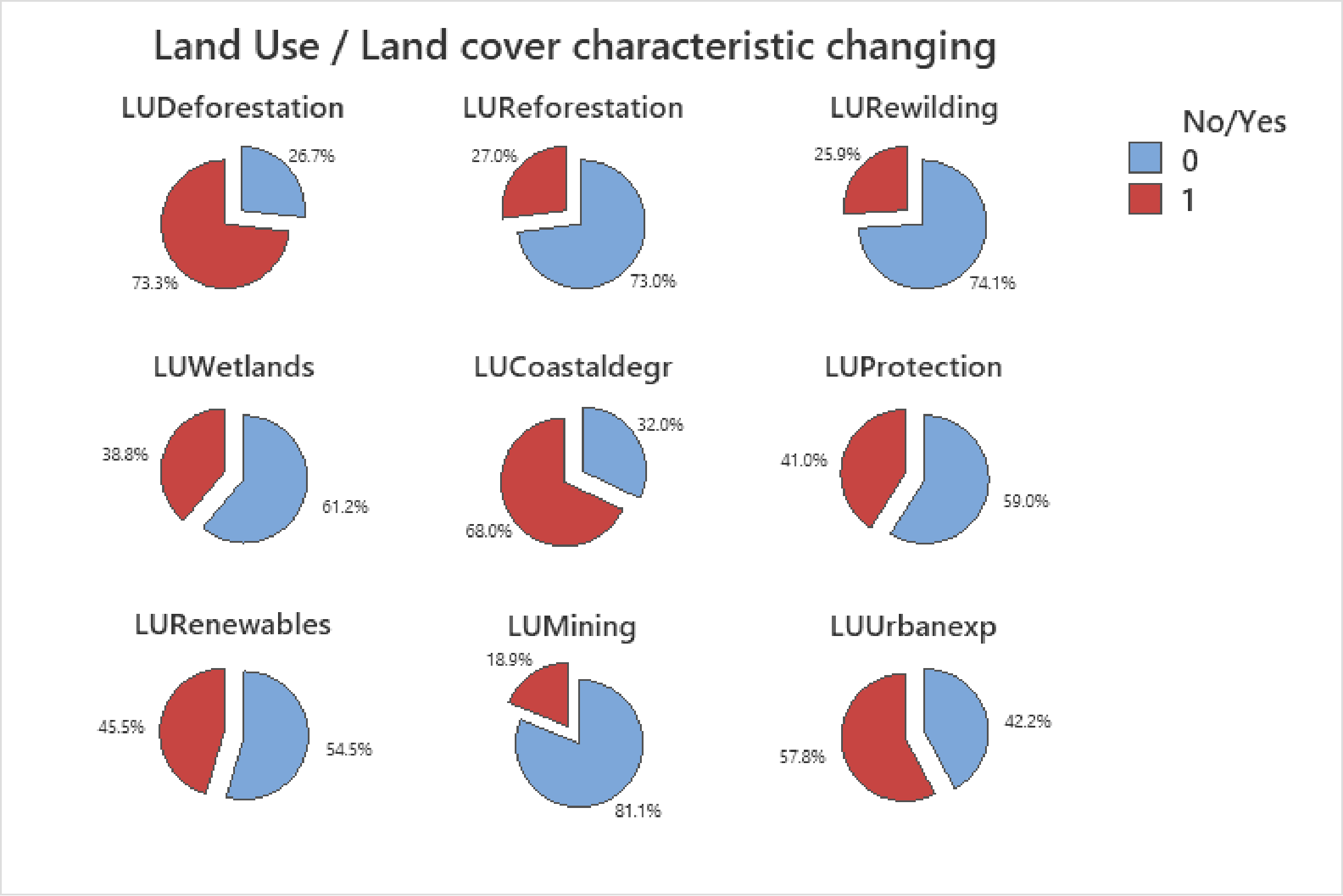

b

*Figure 2.* ***a.*** *Climate variables and reciprocal stakeholder perception of their change.* ***b.*** *Land use variables and reciprocal stakeholder perception of their change. In both cases answers were binary (Yes=1 or No=0) for each variable and allowed for multiple variables that may be changing.*

*Climate change and land use change problems*

Stakeholders ranked water related issues as the most severe problem, with the most common severe score (4) and the highest mean problem score (Fig. 3a, 3b). Primary production sector was the second-highest problem, followed by biodiversity loss (Fig. 3a, 3b). Renewable energy facilities ranked fourth (Fig. 3a, 3b). Urban expansion, habitat destruction, and soil erosion were also significant problems (Fig. 3a, 3b). Infrastructure damage, invasive species, and tourism posed moderate problems, while the lowest problems were associated with carbon sequestration, diseases, pollination, and recreation (Fig. 3a, 3b).

*Overall problems*

Climate change was perceived as the highest overall problem, followed by land-use changes and pollution (Fig. 3c, 3d). Economic growth and austerity ranked next, while nature overexploitation, population growth, and resource extraction were considered minor problems (Fig. 3c, 3d). Policy changes ranked as the lowest overall problem (Fig. 3c, 3d).

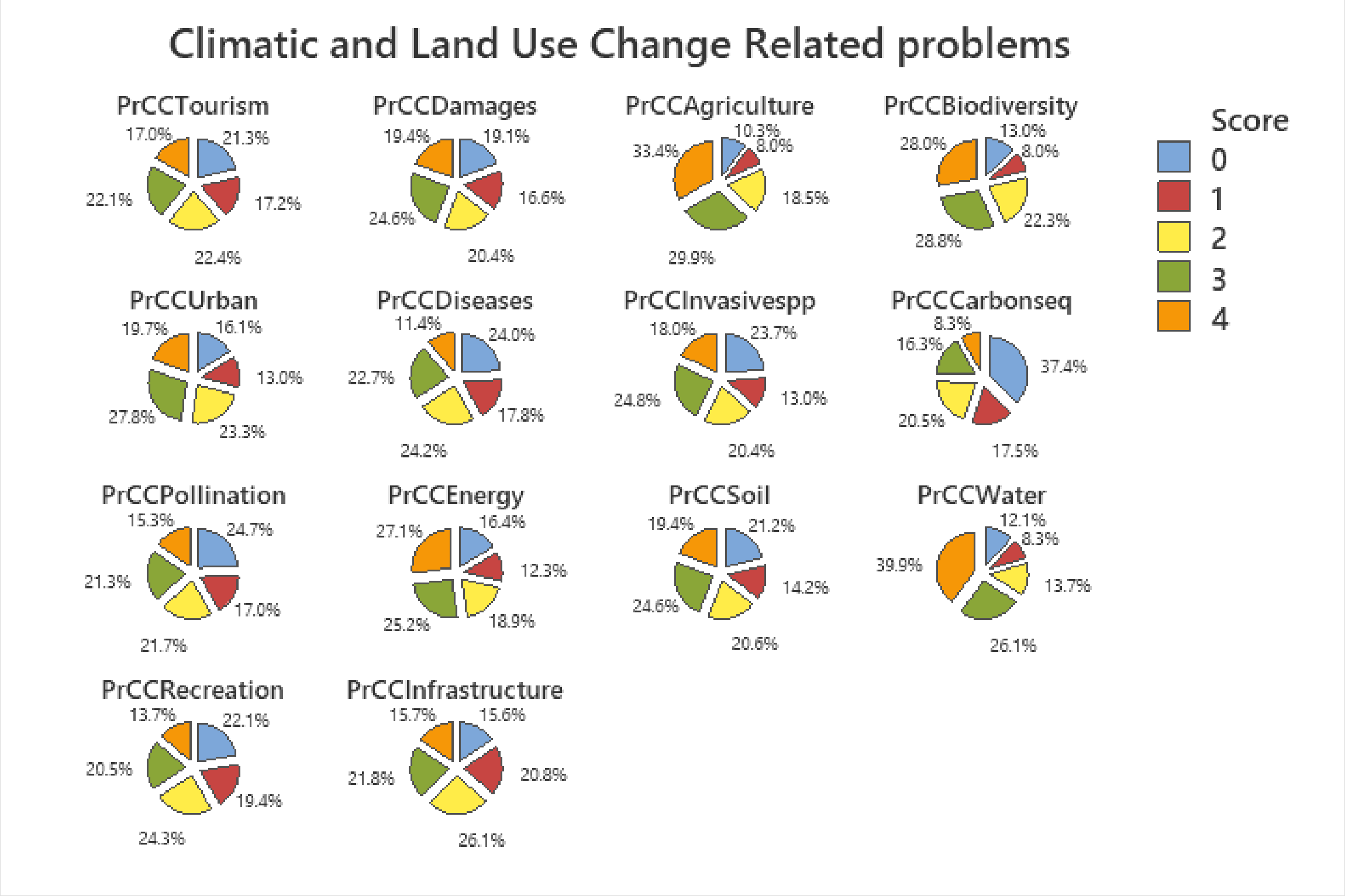

a

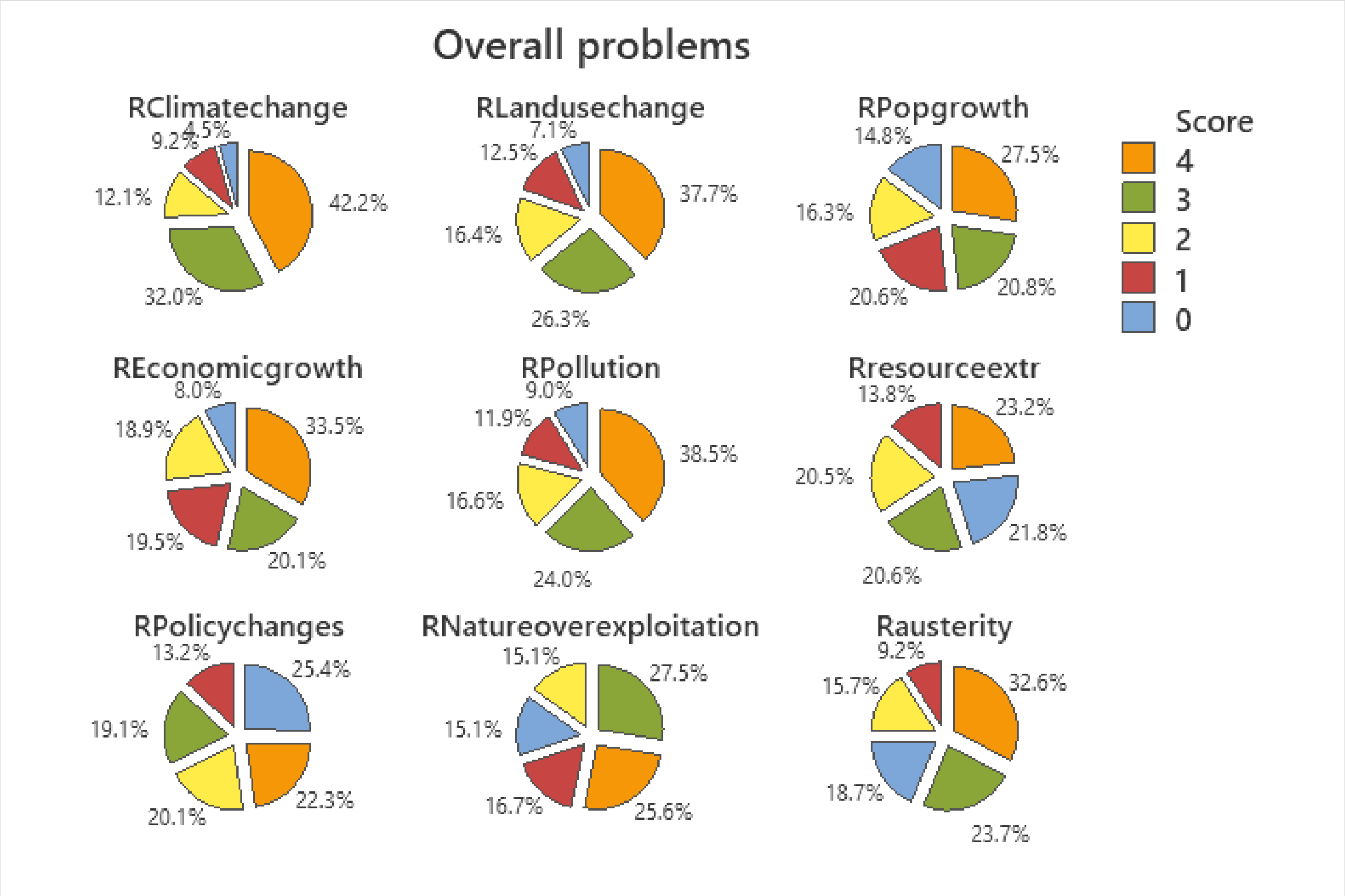

b

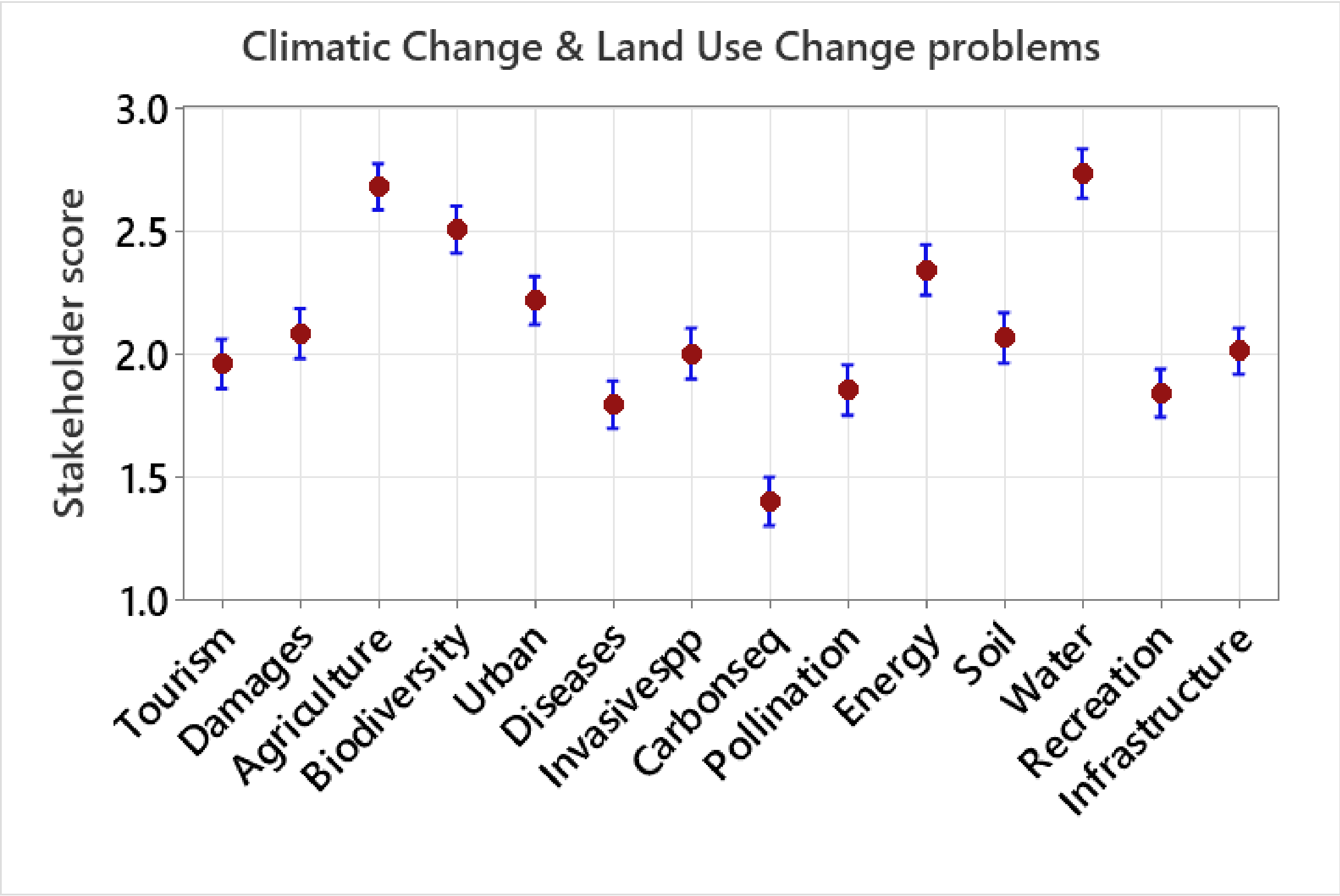

c

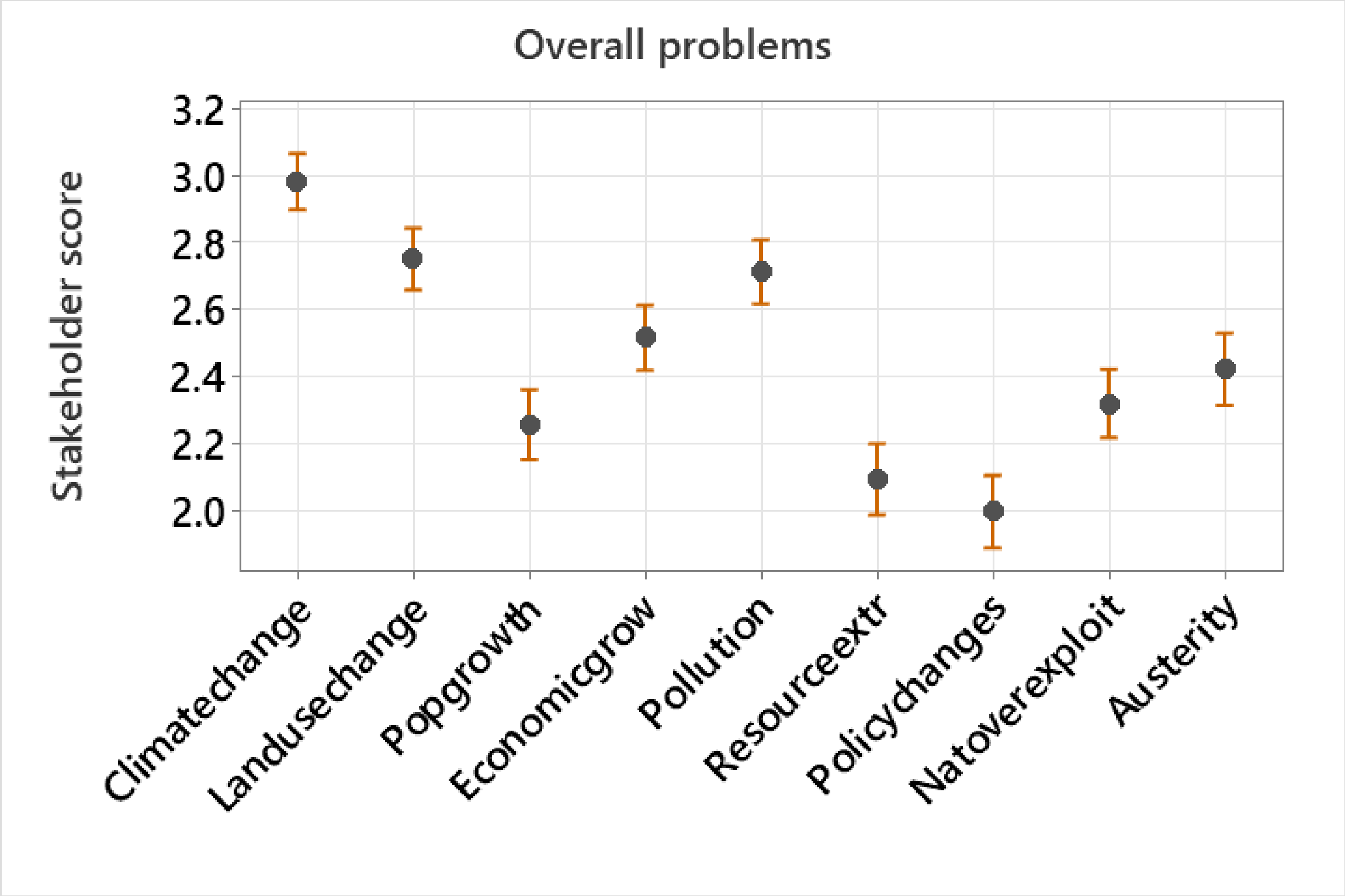

d

***Figure 3.*** *Stakeholder problem perception and problem severity. Higher scores indicate a more severe problem.* ***a.*** *Climate change and land use change combined problem severity ranking.* ***b.*** *Overall problem combining climate change, land use change, other environmental, and social or economic factors*

*ranking.* ***c.*** *Mean score of climate change and land use change combined problem severity indicated in panel a.* ***d.*** *Mean score of overall combined problem severity indicated in panel b.*

*Climate change and Land use change impacts on Ecosystem Services*

Responses showed that 62.8% of stakeholders perceived climate change impacts on ecosystem services as negative, 19.4% as unknown, 14.4% as neutral, and 3.4% as positive impacts (Fig. 4a). For land-use changes, 54.4% indicated negative impacts, 21.4% as unknown, 21.4% as neutral, and 2.7% as positive impacts (Fig. 4b).

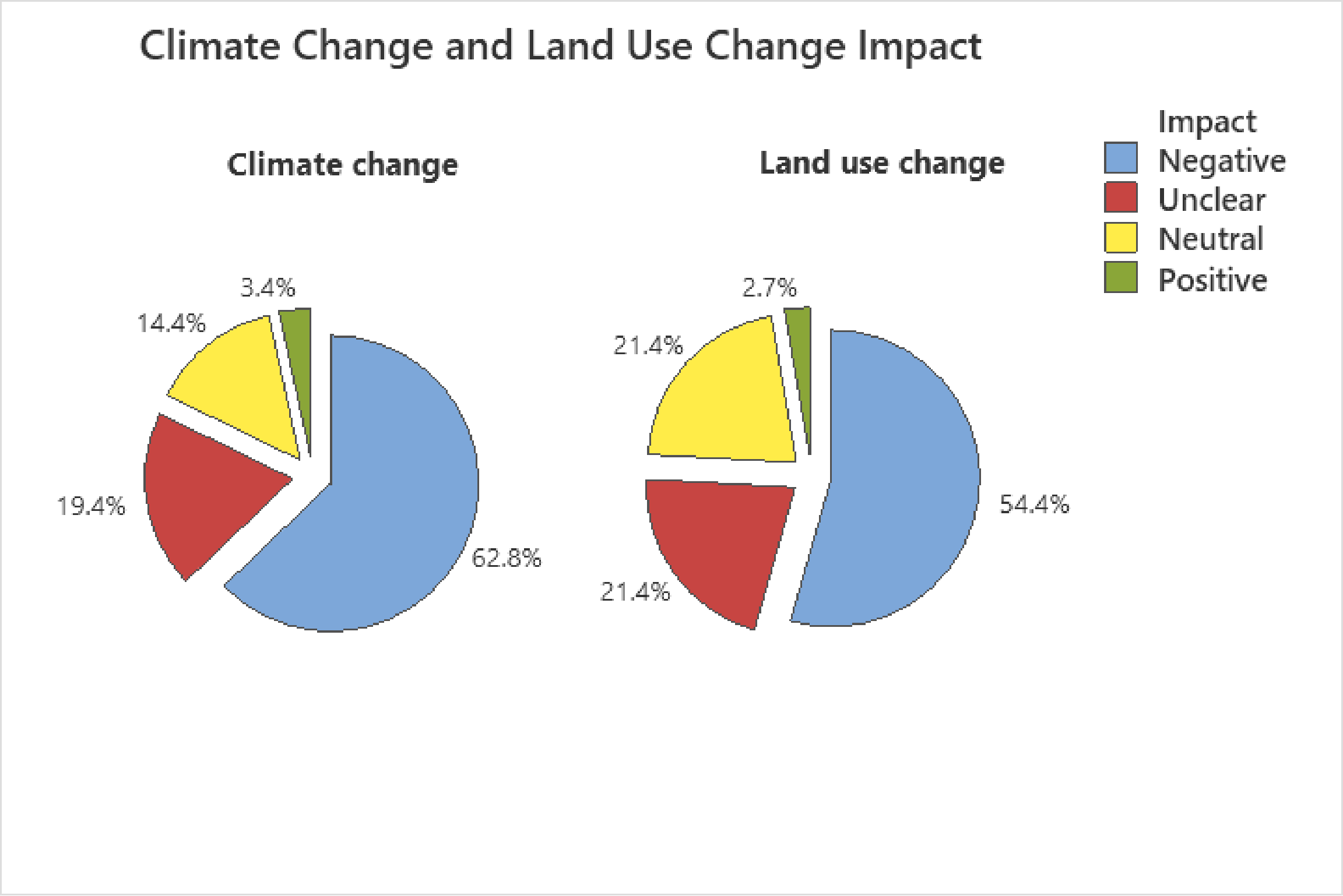

a

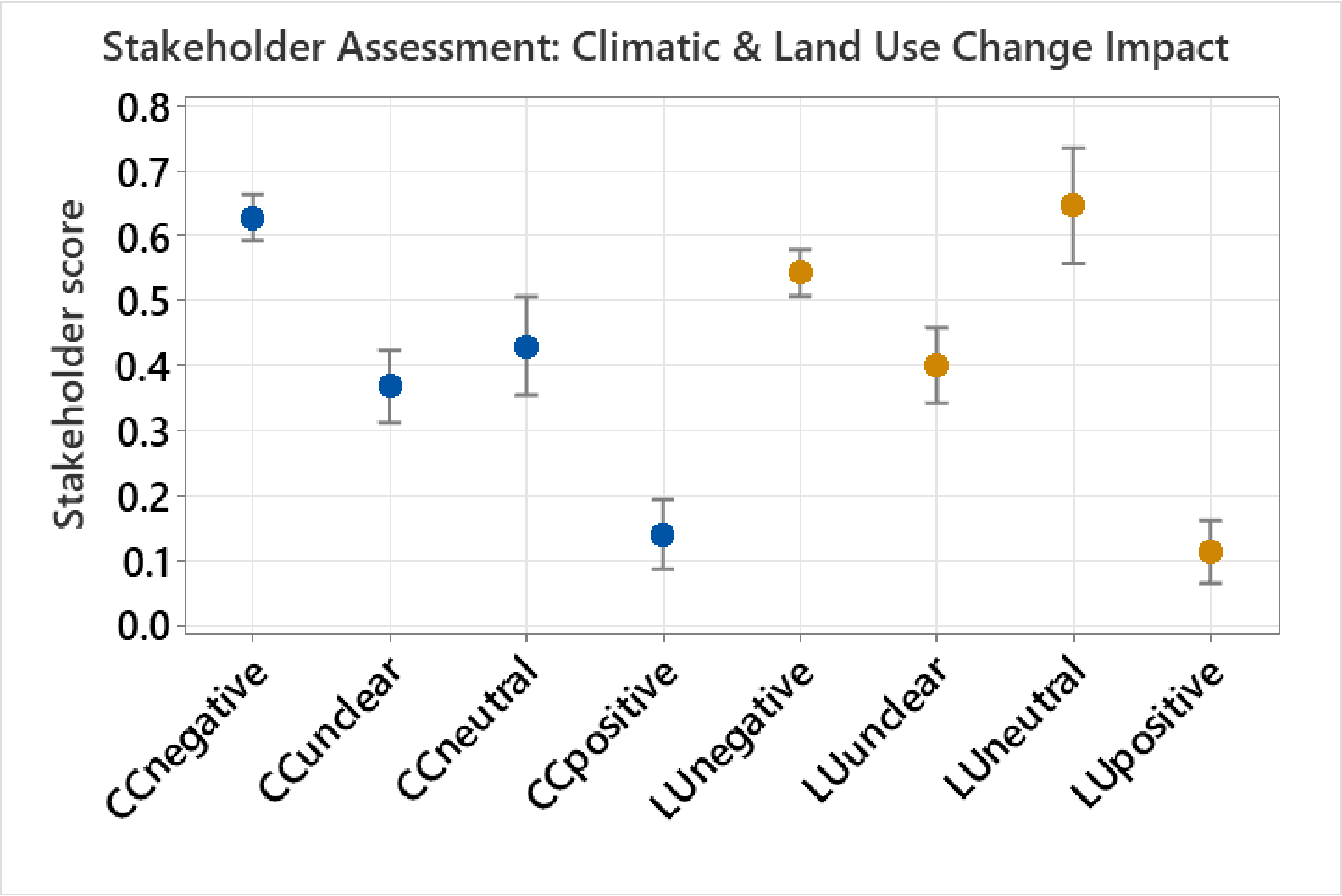

b

***Figure 4.*** *Climate change and land use change impacts on ecosystem services. Impact perceptions were quantified as negative, unknown, neutral, or positive.* ***a.*** *Climate change and land use change impact assessment ranking.* ***b.*** *Mean score of climate change and land use change impact assessment indicated in panel a. Score derived by assigning values of negative = 1, unknown = 2, neutral = 3, and positive = 4.*

*Climate change impact analysis*

Machine learning analysis showed that stakeholders associated negative climate change impacts on ecosystem services primarily with habitat destruction and biodiversity loss, followed by invasive species, overexploitation, water issues, and urban expansion. Agriculture, soil degradation, pollination, and coastal degradation were also significant factors. Classification accuracy was 66.8% (Table 2a). Stakeholders that reported unclear climate change impacts on the benefits of biodiversity were most linked to biodiversity loss, habitat destruction, and water related issues. Additional factors included carbon sequestration, recreation, invasive species, agriculture, urban expansion, soil degradation, and energy-related problems, with a classification accuracy of 61.9% (Fig. 5b, Table 2b). Neutral perceptions of climate change impacts on the benefits of biodiversity were primarily tied to overexploitation and invasive species, followed by carbon sequestration, pollination, resource extraction, and pollution. Soil degradation, austerity, policy changes, and biodiversity loss were also contributing factors. The classification accuracy for neutral impacts was 59.3% (Fig. 5c, Table 2c). There were insufficient stakeholder responses to derive meaningful model coefficients for positive climate change impacts.

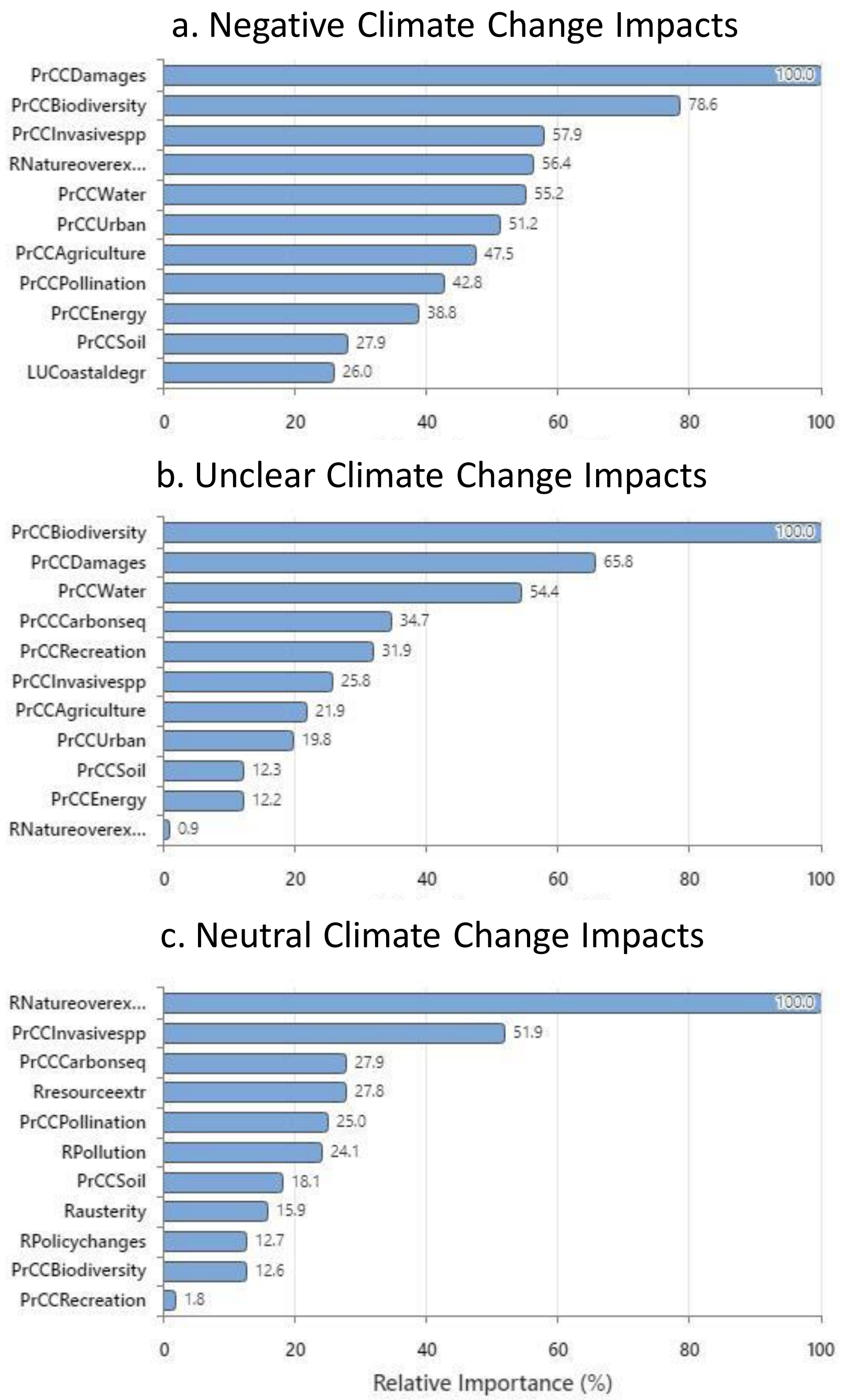

***Figure 5.*** *Relative importance of Climate change impact assessment by island stakeholders quantified by Random Forests machine learning analysis. a. Variable importance of negative climate change impacts stakeholder perceptions. b. Variable importance of unclear climate change impacts*

*stakeholder perceptions. c. Variable importance of neutral climate change impacts stakeholder perceptions.*

*Land use change impact analysis*

Machine learning analysis showed that stakeholders associated negative impacts of land use changes on ecosystem services mainly with coastal degradation, biodiversity loss, pollution, and infrastructure damage. Other contributors included invasive species, economic growth, deforestation, and recreation-related issues, alongside resource extraction, financial austerity, overexploitation, and water scarcity. The classification accuracy for these findings was 62.8% (Fig. 6a, Table 2d). Unclear stakeholder perceptions of land use change impacts on the benefits of biodiversity were most associated with rewilding, urban expansion, and austerity, followed by recreation, policy changes, and wetland problems. The model achieved a classification accuracy of 68.6% (Fig. 6b, Table 2e). Neutral land use change impacts were primarily linked to pollution, overexploitation, and resource extraction, with coastal degradation, invasive species, and diseases also playing significant roles. Factors like water, economic growth, rewilding, and soil-related problems contributed less but reached up to 20% importance. The classification accuracy for neutral impacts was 71.7% (Fig. 6c, Table 2f). Due to limited data entries, the analysis could not effectively model positive land use change impacts.

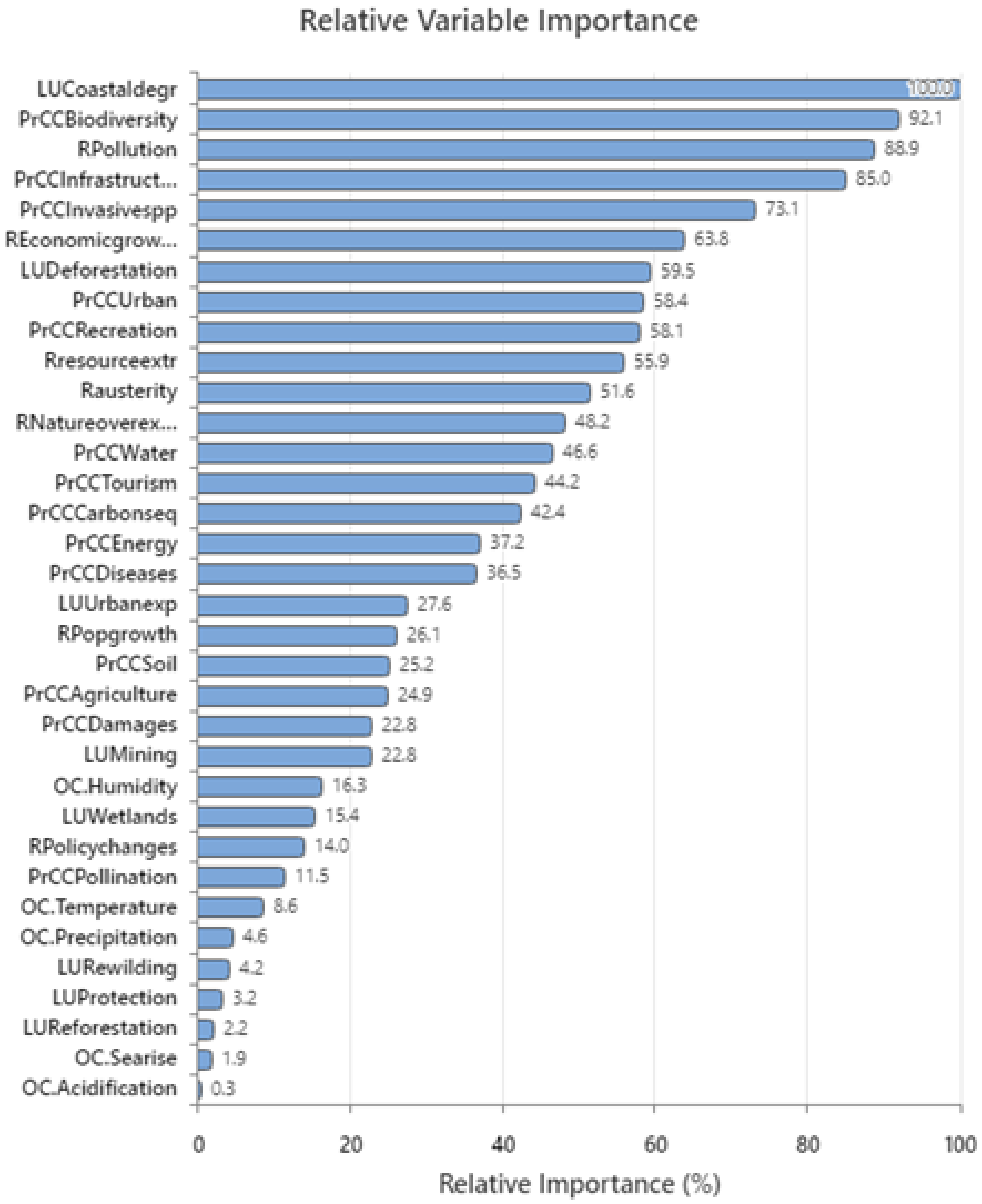

a

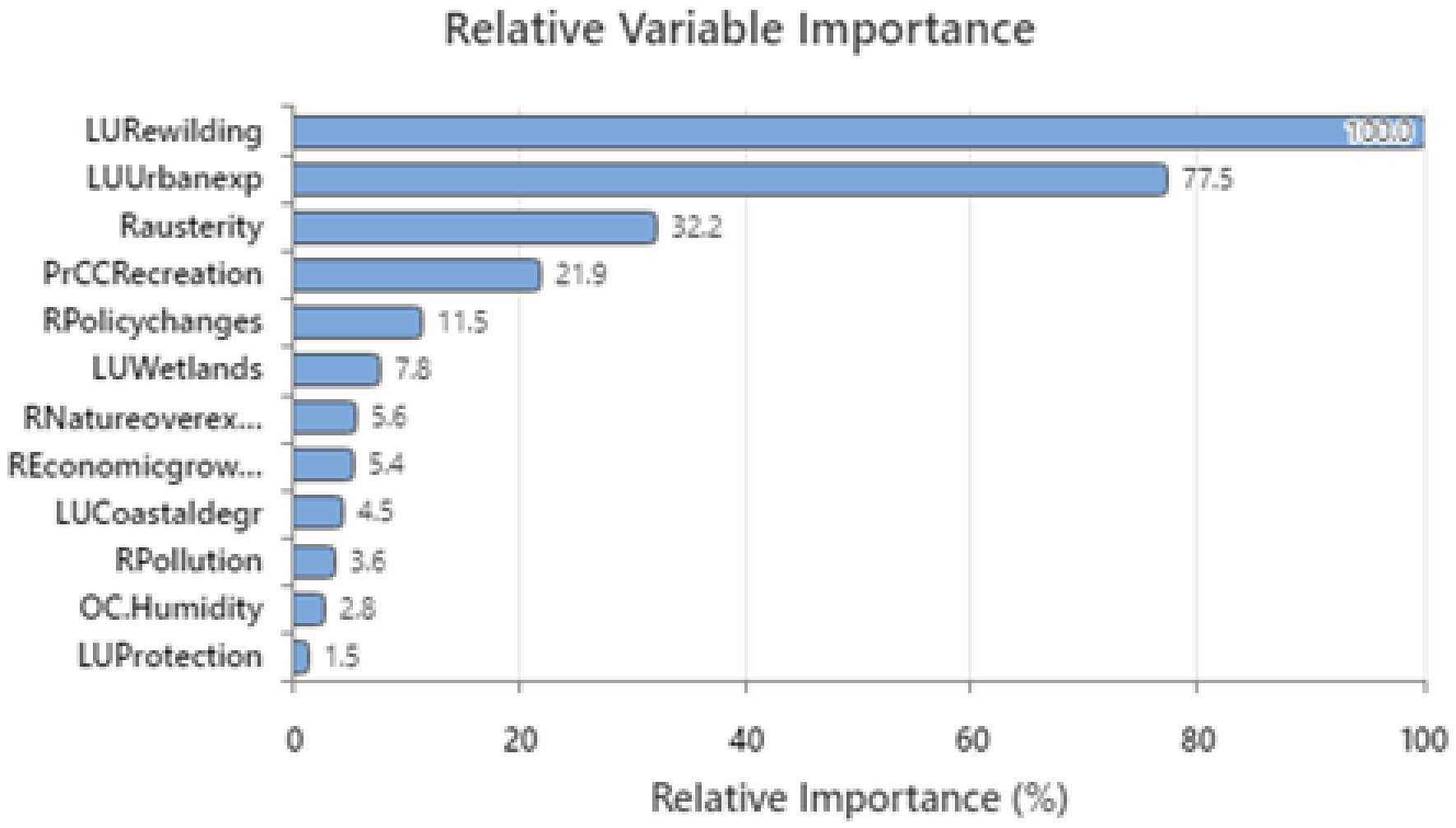

b

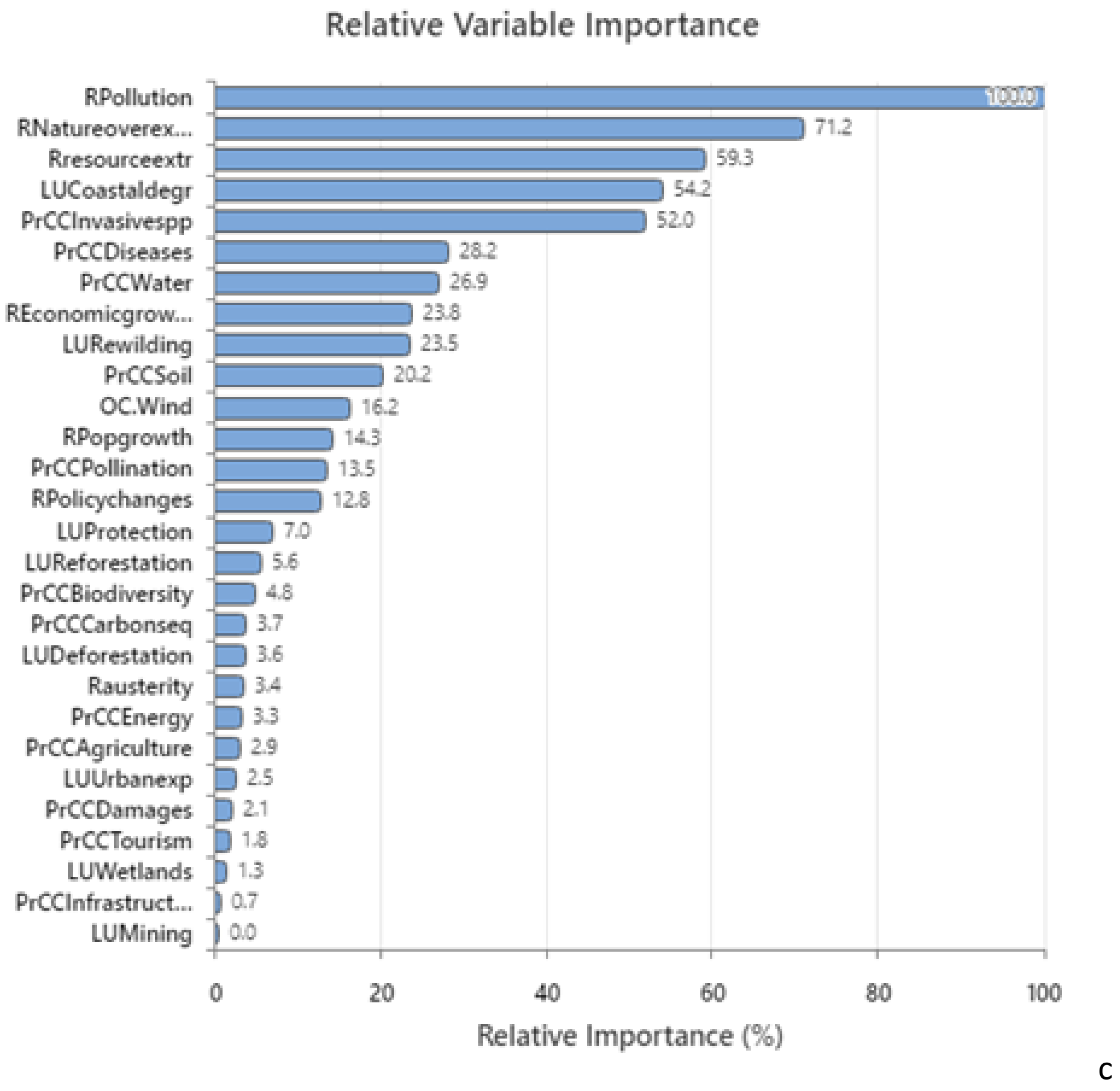

***Figure 6.*** *Relative importance of Land Use change impact assessment by island stakeholders quantified by Random Forests machine learning analysis. a. Variable importance of negative climate change impacts stakeholder perceptions. b. Variable importance of unclear climate change impacts stakeholder perceptions. c. Variable importance of neutral climate change impacts stakeholder perceptions.*

**Table 2.** *Model accuracy in the training (70%) and testing (30%) of the data for (a) negative, (b) unclear, and (c) neutral impacts of climate change on ecosystem services respectively. Accuracy is presented using the model outputs against the testing data not used for model training. Panels d, e, and f indicate model accuracy in the training (70%) and testing (30%) of the data for negative, unclear, and neutral land use change impacts on ecosystem services respectively.*

## Confusion Matrix

| | Predicted Class (Training) | | | | Predicted Class (Test) | | | |
|---|---|---|---|---|---|---|---|---|
| Actual Class | Count | 1 | 0 | % Correct | Count | 1 | 0 | % Correct |
| 1 (Event) | 317 | 197 | 120 | 62.1 | 147 | 94 | 53 | 63.9 |
| 0 | 194 | 60 | 134 | 69.1 | 79 | 22 | 57 | 72.2 |
| All | 511 | 257 | 254 | 64.8 | 226 | 116 | 110 | 66.8 |

| Statistics | Training (%) | Test (%) |
|---|---|---|
| True positive rate (sensitivity or power) | 62.1 | 63.9 |
| False positive rate (type I error) | 30.9 | 27.8 |
| False negative rate (type II error) | 37.9 | 36.1 |
| True negative rate (specificity) | 69.1 | 72.2 |

a

## Confusion Matrix

| | Predicted Class (Training) | | | | Predicted Class (Test) | | | |
|---|---|---|---|---|---|---|---|---|
| Actual Class | Count | 2 | 0 | % Correct | Count | 2 | 0 | % Correct |
| 2 (Event) | 94 | 60 | 34 | 63.8 | 42 | 27 | 15 | 64.3 |
| 0 | 417 | 161 | 256 | 61.4 | 184 | 71 | 113 | 61.4 |
| All | 511 | 221 | 290 | 61.8 | 226 | 98 | 128 | 61.9 |

| Statistics | Training (%) | Test (%) |
|---|---|---|
| True positive rate (sensitivity or power) | 63.8 | 64.3 |
| False positive rate (type I error) | 38.6 | 38.6 |
| False negative rate (type II error) | 36.2 | 35.7 |
| True negative rate (specificity) | 61.4 | 61.4 |

b

## Confusion Matrix

| | Predicted Class (Training) | | | | Predicted Class (Test) | | | |
|---|---|---|---|---|---|---|---|---|
| Actual Class | Count | 3 | 0 | % Correct | Count | 3 | 0 | % Correct |
| 3 (Event) | 76 | 59 | 17 | 77.6 | 30 | 22 | 8 | 73.3 |
| 0 | 435 | 180 | 255 | 58.6 | 196 | 84 | 112 | 57.1 |
| All | 511 | 239 | 272 | 61.4 | 226 | 106 | 120 | 59.3 |

| Statistics | Training (%) | Test (%) |
|---|---|---|
| True positive rate (sensitivity or power) | 77.6 | 73.3 |
| False positive rate (type I error) | 41.4 | 42.9 |
| False negative rate (type II error) | 22.4 | 26.7 |
| True negative rate (specificity) | 58.6 | 57.1 |

c

## Confusion Matrix

| | Predicted Class (Training) | | | | Predicted Class (Test) | | | |
|---|---|---|---|---|---|---|---|---|
| Actual Class | Count | 1 | 0 | % Correct | Count | 1 | 0 | % Correct |
| 1 (Event) | 268 | 213 | 55 | 79.5 | 133 | 88 | 45 | 66.2 |
| 0 | 243 | 46 | 197 | 81.1 | 93 | 39 | 54 | 58.1 |
| All | 511 | 259 | 252 | 80.2 | 226 | 127 | 99 | 62.8 |

| Statistics | Training (%) | Test (%) |
|---|---|---|
| True positive rate (sensitivity or power) | 79.5 | 66.2 |
| False positive rate (type I error) | 18.9 | 41.9 |
| False negative rate (type II error) | 20.5 | 33.8 |
| True negative rate (specificity) | 81.1 | 58.1 |

d

## Confusion Matrix

| | Predicted Class (Training) | | | | Predicted Class (Test) | | | |
|---|---|---|---|---|---|---|---|---|
| Actual Class | Count | 2 | 0 | % Correct | Count | 2 | 0 | % Correct |
| 2 (Event) | 105 | 60 | 45 | 57.1 | 43 | 25 | 18 | 58.1 |
| 0 | 406 | 122 | 284 | 70.0 | 183 | 53 | 130 | 71.0 |
| All | 511 | 182 | 329 | 67.3 | 226 | 78 | 148 | 68.6 |

| Statistics | Training (%) | Test (%) |
|---|---|---|
| True positive rate (sensitivity or power) | 57.1 | 58.1 |
| False positive rate (type I error) | 30.0 | 29.0 |
| False negative rate (type II error) | 42.9 | 41.9 |
| True negative rate (specificity) | 70.0 | 71.0 |

e

## Confusion Matrix

| | Predicted Class (Training) | | | | Predicted Class (Test) | | | |
|---|---|---|---|---|---|---|---|---|
| Actual Class | Count | 3 | 0 | % Correct | Count | 3 | 0 | % Correct |
| 3 (Event) | 117 | 89 | 28 | 76.1 | 42 | 26 | 16 | 61.9 |
| 0 | 394 | 104 | 290 | 73.6 | 184 | 48 | 136 | 73.9 |
| All | 511 | 193 | 318 | 74.2 | 226 | 74 | 152 | 71.7 |

| Statistics | Training (%) | Test (%) |
|---|---|---|
| True positive rate (sensitivity or power) | 76.1 | 61.9 |
| False positive rate (type I error) | 26.4 | 26.1 |
| False negative rate (type II error) | 23.9 | 38.1 |
| True negative rate (specificity) | 73.6 | 73.9 |

f

## Discussion

*Take home findings*

Stakeholders perceive temperature change as the most pressing climatic issue and deforestation as the predominant land use change, highlighting the direct ways in which climate and human activities are altering island environments. Water-related problems were consistently ranked as top priorities, reflecting the central role of freshwater resources for island communities and the sensitivity of water systems to both climatic and land use pressures. Energy concerns, including both supply deficiencies and the challenges posed by renewable energy facilities such as wind and solar, were also identified as significant problems, pointing to the interconnectedness of climate, land use, and energy security in shaping stakeholder perceptions.

Stakeholders generally viewed the impacts of climate change on ecosystem services as negative, particularly through biodiversity loss and natural habitat destruction. Land use change impacts were also seen as negative but appeared to be shaped by a wider range of explanatory factors, reflecting the more complex drivers of land cover transformation. Despite geographical differences among islands, perceptions were broadly consistent, suggesting that island stakeholders share common concerns regarding the benefits biodiversity provides. At the same time, the distinction made between climate- and land use-driven impacts demonstrates that stakeholders are able to differentiate between these pressures, emphasizing the importance of addressing them through both targeted and integrated management measures.

*Temperature*

Results indicate that temperature is the climate characteristic that is perceived most affected by climate change, followed by precipitation. Island stakeholders often perceive temperature and precipitation as the most affected climate characteristic by climate change, in agreement with other island studies (Alcantara *et al.*, 2023) as well as in the mainland (Rankoana 2018). Altered rainfall patterns are also quantified as the second changing climate characteristics in other island studies (Alcantara *et al.*, 2023). Sea-level rise is a concern, but its frequency is comparably lower in European islands. Perceptions of islanders regarding climate change impacts can be influenced by factors such as occupation, environmental engagement, and access to information about climate change (Assis *et al.*, 2023a). Perceptions regarding climate change via temperature changes ranks higher than precipitation in spite of water related problems that are linked with precipitation (Koutroulis et al., 2013). Warmer temperatures are recorded amongst several European islands and thus climate change is more expressed via temperature related issues (Meco *et al.*, 2002; Pla-Rabes *et al.*, 2024; Zittis *et al.*, 2025). It is also important to note that climate perceptions of stakeholders residing on islands do not necessarily coincide with actual data and scientific evidence (Assis *et al.*, 2023b). Perceptions of traditional populations may focus on smaller temporal and spatial scale visions (Assis *et al.*, 2023b). Potential ways to account for temperature-related problems includes nature-based solutions (De Montis *et al.*, 2025).

*De- and reforestation*

Deforestation was the most indicated land use change related factor in islands. Lack of reforestation was also highlighted. Deforestation in European islands is a result of a combination of natural and human activities, including agricultural expansion, urban development, tourism growth, and logging practices (Marathianou *et al.*, 2000; Morales *et al.*, 2009; Kefalas *et al.*, 2019). The growth of tourism has led to the development of new roads, hotels, and recreational areas, encroaching on natural landscapes (McElroy, 2003; Otto *et al.*, 2007). Agricultural activities, such as crop cultivation and livestock raising, often lead to forest clearing, reducing biodiversity and disrupting ecosystems. Logging, both legal and illegal, also contributes to deforestation rates (Marathianou *et al.*, 2000; Morales *et al.*, 2009; Kefalas *et al.*, 2019). Climate change, including rising temperatures and erratic rainfall patterns, weakens forests, making them more vulnerable to pests and fires. In Mediterranean islands like Sardinia and Crete, wildfires are becoming more frequent and devastating, threatening wildlife habitats and local communities (Bacciu *et al.*, 2021). Solutions include protecting remaining forests, establishing nature reserves, national parks, and protected zones, and promoting sustainable tourism through legislation and reforestation projects. Balancing development with conservation is crucial to prevent long-term damage and maintain ecological health (Connell, 2018). In water-scarce and often shallow soil European islands, there is potential for planting or maintaining trees which have also a significant role as biocultural heritage (Camarda and Brundu 2021). In addition the EU Biodiversity Strategy for 2030 favours reforestation (EuropeanCommission 2024).

*Water*

Water-related challenges are topping the list of stakeholders as a problem related with climate and land use changes. Interestingly, in the machine learning analysis accounting for all factors together, it is not topping the list of variables with high explanatory power in explaining climate change or land use change impacts. Thus, when seen together with all other factors, our analysis indicates that stakeholders may perceive water-related problems rather as a matter of bad management (Atay and Saladié 2022) than as a climate change driven impact. Indeed, in the Mediterranean at least, long term precipitation records indicate that precipitation variability is the rule and no trend is identified (Vicente-Serrano *et al.*, 2025). In general problems related to water are often severe, both in terms of scarcity or extreme precipitation events and public interest is high in general (Santín et al., 2023) and in islands (Ricart *et al.*, 2024) but how it is determined by climate change is less clear.

Water is a precious resource, and addressing water-related issues is crucial for a sustainable future. A cross-scale water management system is needed spanning from the island to the local and individual unit level, to address the future water shortages (Atay and Saladié 2022; Hophmayer-Tokich and Kadiman 2006). Authorities should develop water management plans using desalination units, waste recycling, underground resources, and rainwater collection (Angelakis *et al.*, 2025; Cuka, 2025). Agriculture heavily uses water resources as well as greenhouse gases (Yu et al., 2025), and climate-smart agriculture using drip irrigation can reduce freshwater demand (Kourgialas 2021). Revisiting ethical and legal requirements for swimming pools and launching information campaigns to address tourists' higher water volume needs is crucial (Hof & Schmitt, 2011; Doménech-Sánchez *et al.*, 2021). Incorporating water related problems of islands in the EU Water Framework Directive is key to a consistent improvement.

*Energy*

Energy-related problems ranked fourth in terms of combined climate change and land use change problems, while wind and solar energy facilities also ranked as the fourth most important land use problem; nearly half of the stakeholders indicated that the conversion of agricultural land to renewable energy facilities are a land use change problem on their island. Indeed, wind energy facilities induce land take and habitat fragmentation (Kati et al., 2021). In addition, they can pose a threat to biodiversity by inducing avian species loss that are important for the ecosystem functioning (Rebolo-Ifrán *et al.*, 2025). This problem may be more pronounced in islands that their area is more restricted than mainland locations and islands are often part of migration routes providing food and shelter (Assandri et al., 2024). There are several societal barriers on the development of renewable energy facilities on islands (Oikonomou *et al.*, 2009). Island communities may prioritize local benefits and credible mechanisms for managing intra-community conflicts in engagement processes regarding renewable energy facilities (Kallis *et al.*, 2021).

The European Commission's "renewable islands for 2030" initiative has confirmed 30 islands and island groups aiming for energy independence by 2030 using wind and solar facilities. The islands will move towards their own path to climate-neutrality and receive comprehensive assistance. However, spatial planning is often not conducted locally or regionally, stakeholders are insufficiently consulted, and investors decide locations based on investment criteria (Chassot et al., 2014; Kirkegaard et al., 2023) without sufficient environmental risk assessment (Cavallaro & Ciraolo, 2005). Our results show that despite energy being an issue of concern for island stakeholders, current implementation of renewables generates a problem of negative environmental impacts. We are thus facing the two crises, biodiversity loss and climate change, as two separate crises (Pörtner et al., 2023). Climate change is a major driver of biodiversity change, but ecosystem destruction undermines nature's ability to regulate emissions and protect against extreme weather (Seddon 2022). Islands contribute proportionately less to climate change and carbon emissions than reciprocal mainland areas (Nurse *et al.*, 2014). It is thus paradoxical to target complete energy autonomy derived by wind and solar energy, by the ones that actually contribute less to carbon emissions but have limited space and are more fragile to environmental changes. Policies must address these issues together with a stakeholder engagement and consensus (Kallis et al. 2021).

*Impact assessment*

The majority of stakeholders indicated negative climate change impacts on island ecosystem services, followed by unclear and neutral impacts. Very few stakeholders indicated positive climate change impacts on ecosystem services. There is a consensus regarding the main problems of climate change on ecosystem services for stakeholders with negative or unclear impacts views, ranking natural habitat destruction and biodiversity loss as the top problems. Stakeholders with negative perceptions of land use change impacts on ecosystem services identified also biodiversity loss as the main problem. Profiling of stakeholders with neutral climate change impacts on ecosystem services indicates that they believe that it is rather human and other environmental actions impacting ecosystem services via nature over-exploitation, invasive alien species, insufficient carbon sequestration, resources extraction (Tebboth et al., 2020) and not climate change directly. In general, neutral climate change or land use change impact perceptions are poorly linked with biodiversity (Santos et al., 2021; Watt 2020). Land use change unclear or neutral impact perceptions are generally related with coastal degradation and pollution. Coasts are the most impacted environments, especially due to tourism growth and consequent urbanization and the main plant extinctions are reported around the coast (Fois et al., 2018). Notably policy is not a highly ranked variable regarding climate change or land use change impact, indicating that stakeholders are either unaware of related policies or that their opinion is that they have a low overall influence on the output. This can derive either by a lack of trust in policy makers (Kulin and Johansson Sevä 2019) or environmental policies are perceived as paying more in prices and taxes rather than enacting pro-environmental laws despite being agreeable to accepting cuts on living standards (Yousefpour et al., 2020). Nevertheless, land use change policies are critical for the conversion of natural and agricultural land to artificial land cover (Guo et al., 2023; Zeng et al., 2025)

Land use change impacts are more complex to explain than climate change impacts as a higher number of explanatory variables is associated with the impact outcome. Both climate change and land use change impacts are just about equally predictable in terms of model predictive accuracy of the impact in all cases around 60% to 70%. Given the large number of stakeholders, islands, geographic disparity and associated variance, it appears that stakeholders have sufficiently common perceptions regarding climate change and land use change impacts on biodiversity to achieve this explanatory power. In addition, stakeholders substantially differentiate between factors related with climate change impacts and land use change impacts. Globally land use changes have been associated with several-fold higher negative impacts to biodiversity (IPBES 2019; WWF 2020; 2022). In the islands examined here stakeholders expressed the view that climate change has overall more negative impact than land use changes. A worldwide review on impacts of climatic changes and land use changes on ecosystem services indicated that land use changes have more pronounced effects on ecosystem services and that their synergies are key for understanding the impact (Moustakas et al. 2025). This study may not differentiate between perceptions of problems or impacts and facts - i.e. are climate change impacts on biodiversity more pronounced in the study islands than land use changes? Land use changes are in general noteworthy less linked with biodiversity than climate change in the views of stakeholders. However, the most profoundly identified climate change negative impact, habitat destruction, is resulting in a land use change.

*Limitations and future research*

Factors such as the location of the island may have significant effects on the climate change or land use change problem perception and impact as they are correlated with climate and weather patterns (Nambima et al., 2024). In addition, island size may well play a role both on the impacts and problems (Vogiatzakis et al., 2016) together with access to resources to deal with challenges. Stakeholder's profession and reciprocal social or financial status may also influence the way they respond or prioritize locally observable challenges (Sivonen and Koivula 2024). The ecosystem type (marine, terrestrial, freshwater) that the stakeholder is professionally or personally more related to, may also be impacted differently (Scholze et al., 2006). Ultimately mapping perceptions of problems and impacts based on the island, country, or geographic location may provide novel insights into initially quantifying and understanding climate and land use change perceptions and impacts (Palla et al., 2024) as experienced by residents. Stakeholders' perceptions regarding policy efficacy spanning from local, regional, national, and European level can provide novel insights in communicating or adapting policy to the citizens' needs and lived experience of changes (Nguyen et al., 2024). Incorporating open-ended questions into surveys and living labs experience may provide additional insights (Mason and Neumann 2024; Marselis et al., 2024; Song et al., 2021).

**Conclusions**

To the best of our knowledge this is the first study which investigates the interplay of climate changes, land use changes, biodiversity and stakeholders' opinions on islands. This is further pronounced by the geographical scale of the study including a large number of islands under a common policy and legislative framework at a larger scale (i.e. EU administration). Yet these islands experience a different national, regional or local administration, expanding also in seas with very different characteristics (ecological and socio-economic) and different climate change impacts according to future model predictions (Moustakas *et al.*, 2025; Pathirana, 2025). Most of these islands are moreover popular tourist destinations in an era of escalating tourism impact, and thus facing increasing challenges managing their vulnerable natural and cultural ecosystems (Bahçekapılı & Yalçıntan, 2024; Cadima Ribeiro *et al.*, 2025; Leka *et al.*, 2025; Loftsdóttir & Mixa, 2025).

The topic of environmental change, including mainly climate and land use changes, is intricate and varied, with significant effects on biodiversity, economy, and society (WWF 2020; 2022). It can be challenging to increase public awareness and comprehension of its causes and effects. In order to pursue successful climate adaptation or mitigation strategies and promote climate resilience, it is important to understand the perspectives and knowledge gaps of stakeholders involved in functions that are affected by or address land use and climate change (Ruiz et al., 2023). Combining climate change with land use change (Louca et al., 2015) as well as other socioeconomic and environmental factors is critical for assessing the stakeholder perceptions and ranking problems and impacts (Ratnayake et al., 2024).

The majority of stakeholders indicated negative impacts of climate change on island ecosystem services, with natural habitat destruction and biodiversity loss being the top problems. Stakeholders with neutral perceptions believe that human and environmental actions, such as nature over-exploitation and invasive species, are causing ecosystem services to be negatively impacted. Land use change impacts are more complex to explain than climate change impacts, with both being equally predictable in terms of model predictive accuracy. However, land use changes are generally less linked with biodiversity than climate change in the views of stakeholders. The most profoundly identified climate change negative impact, habitat destruction, is resulting from land use change. Policy is not a highly ranked variable regarding climate change or land use change impacts, suggesting that stakeholders may be unaware of related policies or have a low overall influence on the output. Water, energy, and renewable energy related issues pose serious concerns to island stakeholders and management and policy measures are needed to address those issues. Temperature related problems need to be accounted for. Island-specific European policies may be necessary to address these issues.

**Acknowledgements**

Comments from two anonymous reviewers considerably improved an earlier manuscript draft.

## References

Adamides G. A review of climate-smart agriculture applications in Cyprus. Atmosphere 2020;11: 898.

Alif MDN and Fahrudin NF.Performance Analysis of Oversampling and Undersampling on Telco Churn Data Using Naive Bayes, SVM And Random Forest Methods. E3S Web of Conferences: EDP Sciences; 2024; 02004.

Antronico L, Coscarelli R, De Pascale F, Di Matteo D. Climate Change and Social Perception: A Case Study in Southern Italy. Sustainability 2020;12: 6985.

Assandri G, Bazzi G, Bermejo-Bermejo A, Bounas A, Calvario E, Catoni C, Catry I, Catry T, Champagnon J, De Pascalis F. Assessing exposure to wind turbines of a migratory raptor through its annual life cycle across continents. Biol Conserv 2024;293: 110592.

Atay I and Saladié Ò. Water Scarcity and Climate Change in Mykonos (Greece): The Perceptions of the Hospitality Stakeholders. Tourism and Hospitality 2022;3: 765-787.

Authier R, Pillot B, Guimbretière G, Corral-Broto P, Gervet C. Towards sustainable land management in small islands: A Water-Energy-Food nexus approach. PLOS ONE 2024;19: e0310632.

Balzter H, Macul M, Delaney B, Tansey K, Espirito-Santo F, Ofoegbu C, Petrovskii S, Forchtner B, Nicholes N, Payo E. Loss and damage from climate change: knowledge gaps and interdisciplinary approaches. Sustainability 2023;15: 11864.

Becken S and Wilson J. The impacts of weather on tourist travel. Tourism Geographies 2013;15: 620-639.

Becken S. Water equity–Contrasting tourism water use with that of the local community. Water resources and industry 2014;7: 9-22.

Bjarnason A, Katsanevakis S, Galanidis A, Vogiatzakis IN, Moustakas A. Evaluating Hypotheses of Plant Species Invasions on Mediterranean Islands: Inverse Patterns between Alien and Endemic Species. Frontiers in Ecology and Evolution 2017;5: 91.

Blaće A, Cvitanović M, Čuka A, Faričić J. Land Use/Land Cover Changes on Croatian Islands Since the Beginning of the Twentieth Century—Drivers and Consequences. In: Environmental Histories of the Dinaric Karst. Springer; 2024; 141-165.

Brandt M, Tucker CJ, Kariryaa A, Rasmussen K, Abel C, Small J, Chave J, Rasmussen LV, Hiernaux P, Diouf AA, Kergoat L, Mertz O, Igel C, Gieseke F, Schöning J, Li S, Melocik K, Meyer J, Sinno S, Romero E, Glennie E, Montagu A, Dendoncker M, Fensholt R. An unexpectedly large count of trees in the West African Sahara and Sahel. Nature 2020;587: 78-82.

Breiman L. Random forests. Mach Learn 2001;45.

Butler JRA, Skewes T, Mitchell D, Pontio M, Hills T. Stakeholder perceptions of ecosystem service declines in Milne Bay, Papua New Guinea: Is human population a more critical driver than climate change? Marine Policy 2014;46: 1-13.

Camarda I and Brundu G. Monumental trees and old-growth forests in Sardinia (Italy). Fl Medit 2021;31: 407-414.

Chassot S, Hampl N, Wüstenhagen R. When energy policy meets free-market capitalists: The moderating influence of worldviews on risk perception and renewable energy investment decisions. Energy Research & Social Science 2014;3: 143-151.

Clayton S and Swim JK. Climate change impacts on mental health and well-being. In: APA handbook of health psychology, Volume 3: Health psychology and public health, Vol 3. American Psychological Association; 2025; 401-418.

Dale VH, Efroymson RA, Kline KL. The land use–climate change–energy nexus. Landscape Ecol 2011;26: 755-773.

Daliakopoulos IN, Katsanevakis S, Moustakas A. Spatial Downscaling of Alien Species Presences Using Machine Learning. Frontiers in Earth Science 2017;5: 60.
De Chazal J and Rounsevell MD. Land-use and climate change within assessments of biodiversity change: a review. Global Environmental Change 2009;19: 306-315.
Dhar T, Bornstein L, Lizarralde G, Nazimuddin SM. Risk perception—A lens for understanding adaptive behaviour in the age of climate change? Narratives from the Global South. International Journal of Disaster Risk Reduction 2023;95: 103886.
Elmqvist T, Maltby E, Barker T, Mortimer M, Perrings C, Aronson J, De Groot R, Fitter A, Mace G, Norberg J. Biodiversity, ecosystems and ecosystem services. In: The Economics of Ecosystems and Biodiversity: Ecological and economic foundations. Routledge; 2012; 41-111.
EuropeanCommission. Biodiversity strategy for 2030. 2024; Available from: https://environment.ec.europa.eu/strategy/biodiversity-strategy-2030_en.
Falkland T.Water resources issues of small island developing states. Natural resources forum: Wiley Online Library; 1999; 245-260.
Fernández H, Picazo P, Moreno Gil S. The Pathway to Sustainability in a Mass Tourism Destination: The Case of Lanzarote. Sustainability 2024;16: 5253.
Ferrari S, Zagarella F, Caputo P, Beccali M. Mapping seasonal variability of buildings electricity demand profiles in mediterranean small islands. Energies 2023;16: 1568.
Fisher A, Rudin C, Dominici F. All models are wrong, but many are useful: Learning a variable's importance by studying an entire class of prediction models simultaneously. Journal of Machine Learning Research 2019;20: 1-81.
Fisher J, Allen S, Yetman G, Pistolesi L. Assessing the influence of landscape conservation and protected areas on social wellbeing using random forest machine learning. Scientific Reports 2024;14: 11357.
Fois M, Bacchetta G, Cuena-Lombraña A, Cogoni D, Pinna MS, Sulis E, Fenu G. Using extinctions in species distribution models to evaluate and predict threats: a contribution to plant conservation planning on the island of Sardinia. Environmental Conservation 2018;45: 11-19.
Guo X, Zhang Y, Guo D, Lu W, Xu H. How does ecological protection redline policy affect regional land use and ecosystem services? Environmental Impact Assessment Review 2023;100: 107062.
Haines-Young R. Land use and biodiversity relationships. Land Use Policy 2009;26: S178-S186.
Handmer J, Honda Y, Kundzewicz ZW, Arnell N, Benito G, Hatfield J, Mohamed IF, Peduzzi P, Wu S, Sherstyukov B. Changes in impacts of climate extremes: human systems and ecosystems. Managing the risks of extreme events and disasters to advance climate change adaptation special report of the intergovernmental panel on climate change 2012: 231-290.
Hophmayer-Tokich S and Kadiman T.Water management on islands–Common issues and possible actions. Concept paper in preparation to the international workshop: Capacity building in water management for sustainable tourism on islands; 2006.
IPBES.Global assessment report on biodiversity and ecosystem services of the Intergovernmental Science-Policy Platform on Biodiversity and Ecosystem Services. In: ES Brondizio, J Settele, S Díaz, HT Ngo, editors.Bonn, Germany; 2019.
Kati V, Kassara C, Vrontisi Z, Moustakas A. The biodiversity-wind energy-land use nexus in a global biodiversity hotspot. Science of The Total Environment 2021;768: 144471.
Kelman I. Critiques of island sustainability in tourism. In: Island Tourism Sustainability and Resiliency. Routledge; 2022; 36-53.
Kirkegaard JK, Rudolph D, Nyborg S, Cronin T. The landrush of wind energy, its socio-material workings, and its political consequences: On the entanglement of land and wind assemblages in Denmark. Environment and Planning C: Politics and Space 2023;41: 548-566.

Kongsager R. Linking climate change adaptation and mitigation: A review with evidence from the land-use sectors. Land 2018;7: 158.
Kopnina H, Zhang SR, Anthony S, Hassan A, Maroun W. The inclusion of biodiversity into Environmental, Social, and Governance (ESG) framework: A strategic integration of ecocentric extinction accounting. Journal of Environmental Management 2024;351: 119808.
Kourgialas NN. A critical review of water resources in Greece: The key role of agricultural adaptation to climate-water effects. Science of The Total Environment 2021;775: 145857.
Koutroulis AG, Tsanis IK, Daliakopoulos IN, Jacob D. Impact of climate change on water resources status: A case study for Crete Island, Greece. J Hydrology 2013;479: 146-158.
Kuang Y, Zhang Y, Zhou B, Li C, Cao Y, Li L, Zeng L. A review of renewable energy utilization in islands. Renewable and Sustainable Energy Reviews 2016;59: 504-513.
Kulin J and Johansson Sevä I. The role of government in protecting the environment: quality of government and the translation of normative views about government responsibility into spending preferences. International Journal of Sociology 2019;49: 110-129.
Louca M, Vogiatzakis IN, Moustakas A. Modelling the combined effects of land use and climatic changes: Coupling bioclimatic modelling with Markov-chain Cellular Automata in a case study in Cyprus. Ecological Informatics 2015;30: 241-249.
Marselis SM, Hannula SE, Trimbos KB, Berg MP, Bodelier PLE, Declerck SAJ, Erisman JW, Kuramae EE, Nanu A, Veen GF, van 't Zelfde M, Schrama M. The use of living labs to advance agro-ecological theory in the transition towards sustainable land use: A tale of two polders. Environmental Impact Assessment Review 2024;108: 107588.
Martin del Campo F, Singh SJ, Fishman T, Thomas A, Noll D, Drescher M. Can a small island nation build resilience? The significance of resource-use patterns and socio-metabolic risks in The Bahamas. Journal of Industrial Ecology 2023;27: 491-507.
Mason CW and Neumann P. The Impacts of Climate Change on Tourism Operators, Trail Experience and Land Use Management in British Columbia's Backcountry. Land 2024;13: 69.
McEvoy D, Tara A, Vahanvati M, Ho S, Gordon K, Trundle A, Rachman C, Qomariyah Y. Localized nature-based solutions for enhanced climate resilience and community wellbeing in urban informal settlements. Climate and Development 2024;16: 600-612.
Moreno-Alcayde Y, Traver VJ, Leiva LA. Sneaky emotions: impact of data partitions in affective computing experiments with brain-computer interfacing. Biomedical Engineering Letters 2024;14: 103-113.
Moustakas A and Davlias O. Minimal effect of prescribed burning on fire spread rate and intensity in savanna ecosystems. Stoch Environ Res Risk Assess 2021.
Mycoo MA and Roopnarine RR. Water resource sustainability: Challenges, opportunities and research gaps in the English-speaking Caribbean Small Island Developing States. PLOS Water 2024;3: e0000222.
Nambima AB, Houehanou TD, Yehouenou N, Adjacou DM, Alassiri AS, Gouwakinnou G. Socioenvironmental Drivers of Farmers' Perceptions of Climate Change Risk in Agroforestry Parklands of West Atacora in Benin (West Africa). Open Journal of Ecology 2024;14: 54-65.
Newman RJS, Capitani C, Courtney-Mustaphi C, Thorn JPR, Kariuki R, Enns C, Marchant R. Integrating Insights from Social-Ecological Interactions into Sustainable Land Use Change Scenarios for Small Islands in the Western Indian Ocean. Sustainability 2020;12: 1340.
Nguyen M-H, Duong M-PT, Nguyen Q-L, La V-P, Hoang V-Q. In search of value: the intricate impacts of benefit perception, knowledge, and emotion about climate change on marine protection support. Journal of Environmental Studies and Sciences 2024: 1-19.
Nie Y, Avraamidou S, Xiao X, Pistikopoulos EN, Li J, Zeng Y, Song F, Yu J, Zhu M. A Food-Energy-Water Nexus approach for land use optimization. Science of The Total Environment 2019;659: 7-19.

Otto-Banaszak I, Matczak P, Wesseler J, Wechsung F. Different perceptions of adaptation to climate change: a mental model approach applied to the evidence from expert interviews. Regional environmental change 2011;11: 217-228.
Palla A, Pezzagno M, Spadaro I, Ermini R. Participatory Approach to Planning Urban Resilience to Climate Change: Brescia, Genoa, and Matera—Three Case Studies from Italy Compared. Sustainability 2024;16: 2170.
Pichot F, Mouillot D, Juhel JB, Dalongeville A, Adam O, Arnal V, Bockel T, Boulanger E, Boissery P, Cancemi M. Mediterranean Islands as Refugia for Elasmobranch and Threatened Fishes. Diversity and Distributions 2024: e13937.
Pörtner H-O, Scholes R, Arneth A, Barnes D, Burrows MT, Diamond S, Duarte CM, Kiessling W, Leadley P, Managi S. Overcoming the coupled climate and biodiversity crises and their societal impacts. Science 2023;380: eabl4881.
Rankoana SA. Human perception of climate change. Weather 2018;73: 367-370.
Rasmus S, Yletyinen J, Sarkki S, Landauer M, Tuomi M, Arneberg MK, Bjerke JW, Ehrich D, Habeck JO, Horstkotte T, Kivinen S, Komu T, Kumpula T, Leppänen L, Matthes H, Rixen C, Stark S, Sun N, Tømmervik H, Forbes BC, Eronen JT. Policy documents considering biodiversity, land use, and climate in the European Arctic reveal visible, hidden, and imagined nexus approaches. One Earth 2024;7: 265-279.
Ratnayake SS, Reid M, Larder N, Hunter D, Hasan MK, Dharmasena PB, Kogo B, Senavirathna M, Kariyawasam CS. Climate and Land Use Change Pressures on Food Production in Social-Ecological Systems: Perceptions from Farmers in Village Tank Cascade Systems of Sri Lanka. Sustainability 2024;16: 8603.
Ruiz I, Pompeu J, Ruano A, Franco P, Balbi S, Sanz MJ. Combined artificial intelligence, sustainable land management, and stakeholder engagement for integrated landscape management in Mediterranean watersheds. Environmental Science & Policy 2023;145: 217-227.
Sakti AD, Adillah KP, Santoso C, Al Faruqi I, Hendrawan VSA, Sofan P, Fauzi AI, Setiawan Y, Utami I, Zain AF. Modeling Proboscis monkey conservation sites on Borneo using ensemble machine learning. Global Ecology and Conservation 2024;54: e03101.
Salas Reyes R, Nguyen VM, Schott S, Berseth V, Hutchen J, Taylor J, Klenk N. A Research Agenda for Affective Dimensions in Climate Change Risk Perception and Risk Communication. Frontiers in Climate 2021;3.
Santín C, Moustakas A, Doerr SH. Searching the flames: Trends in global and regional public interest in wildfires. Environmental Science & Policy 2023;146: 151-161.
Santos MJ, Smith AB, Dekker SC, Eppinga MB, Leitão PJ, Moreno-Mateos D, Morueta-Holme N, Ruggeri M. The role of land use and land cover change in climate change vulnerability assessments of biodiversity: a systematic review. Landscape Ecol 2021;36: 3367-3382.
Scholze M, Knorr W, Arnell NW, Prentice IC. A climate-change risk analysis for world ecosystems. Proceedings of the National Academy of Sciences 2006;103: 13116-13120.
Seddon N. Harnessing the potential of nature-based solutions for mitigating and adapting to climate change. Science 2022;376: 1410-1416.
Sivonen J and Koivula A. How do social class position and party preference influence support for fossil fuel taxation in Nordic countries? The Social Science Journal 2024;61: 453-473.
Solé Figueras L, Zandt EI, Buschbaum C, Meunier CL. How are the impacts of multiple anthropogenic drivers considered in marine ecosystem service research? A systematic literature review. J Appl Ecol 2024.
Song J, Ko Y, Hwang J, Kim C. Exploring the Feasibility of the Living Lab Approach in Addressing Climate Change. The International Journal of Climate Change: Impacts and Responses 2021;14: 15.

Steibl S, Steiger S, Wegmann AS, Holmes ND, Young HS, Carr P, Russell JC. Atolls are globally important sites for tropical seabirds. Nature Ecology & Evolution 2024;8: 1907-1915.

Tebboth MGL, Few R, Assen M, Degefu MA. Valuing local perspectives on invasive species management: Moving beyond the ecosystem service-disservice dichotomy. Ecosystem Services 2020;42: 101068.

Thaman R. Threats to Pacific Island biodiversity and biodiversity conservation in the Pacific Islands. Development Bulletin 2002;58: 23-27.

Tourlioti PN, Portman ME, Pantelakis I, Tzoraki O. Awareness and willingness to engage in climate change adaptation and mitigation: Results from a survey of Mediterranean islanders (Lesvos, Greece). Climate Services 2024;33: 100427.

van der Geest K, Burkett M, Fitzpatrick J, Stege M, Wheeler B. Climate change, ecosystem services and migration in the Marshall Islands: are they related? Climatic Change 2020;161: 109-127.

Venkateswarlu T and Anmala J. Importance of land use factors in the prediction of water quality of the Upper Green River watershed, Kentucky, USA, using random forest. Environment, Development and Sustainability 2024;26: 23961-23984.

Vogiatzakis I, Mannion A, Sarris D. Mediterranean island biodiversity and climate change: the last 10,000 years and the future. Biodiversity and conservation 2016;25: 2597-2627.

Vogiatzakis IN, Balzan MV, Drakou EG, Katsanevakis S, Padoa-Schioppa E, Tzirkalli E, Zotos S, Álvarez X, Külvik M, Fonseca C, Moustakas A, Martínez-López J, Mackelworth P, Mandzukovski D, Ricci L, Srdjevic B, Tase M, Terkenli TS, Zemah-Shamir S, Zittis G, Manolaki P. Enhancing Small-Medium IsLands resilience by securing the sustainability of Ecosystem Services: the SMILES Cost Action. Research Ideas and Outcomes 2023;9: e116061.

Watt A. Land-Use Intensity and Land-Use Change: Impacts on Biodiversity. In:  Life on Land.Cham: Springer International Publishing; 2020; 1-13.

WWF.Living Planet Report 2020 - Bending the curve of biodiversity loss. Almond, R.E.A., Grooten M. and Petersen, T. (Eds). WWF, Gland, Switzerland.; 2020.

WWF.Living Planet Report 2022 – Building a nature-positive society (eds. Almond REA, Grooten M, Juffe Bignoli D, Petersen T). WWF, Gland, Switzerland.; 2022.

Yousefpour R, Prinz A, Ng C. Public perceptions of climate change adaptation in Singapore dealing with forecasted sea level rise. Human and Ecological Risk Assessment: An International Journal 2020;26: 1449-1475.

Yu B, Liu X, Bi X, Sun H, Buysse J. Agricultural resource management strategies for greenhouse gas mitigation: The land-energy-food-waste nexus based on system dynamics model. Environmental Impact Assessment Review 2025;110: 107647.

Zeng L, Yang L, Su L, Hu H, Feng C. The impact of policies on land use and land cover changes in the Beijing–Tianjin–Hebei region in China. Environmental Impact Assessment Review 2025;110: 107676.

Alcantara, L.B., Creencia, L.A., Madarcos, J.R.V., Madarcos, K.G., Jontila, J.B.S. & Culhane, F. (2023) Climate change awareness and risk perceptions in the coastal marine ecosystem of Palawan, Philippines. *UCL Open Environment*, **5**, e054.
Angelakis, A.N., Zafeirakou, A., Kourgialas, N.N. & Voudouris, K. (2025) The Evolution of Unconventional Water Resources in the Hellenic World. *Sustainability*, **17**, 2388.
Aretano, R., Petrosillo, I., Zaccarelli, N., Semeraro, T. & Zurlini, G. (2013) People perception of landscape change effects on ecosystem services in small Mediterranean islands: A combination of subjective and objective assessments. *Landscape and Urban Planning*, **112**, 63-73.
Assis, D.M.S., Medeiros-Sarmento, P.S., Tavares-Martins, A.C.C. & Godoy, B.S. (2023a) Are perceptions of climate change in Amazonian coastal communities influenced by socioeconomic and cultural factors? *Heliyon*, **9**, e18392.
Assis, D.M.S., Franco, V.S., Dias, T.S.S., Sodré, G.R.C., Tavares-Martins, A.C.C. & Godoy, B.S. (2023b) Local perceptions do not follow rainfall trends: A case study in traditional Marajo island communities (eastern para state, BR). *Heliyon*, **9**
Bacciu, V., Hatzaki, M., Karali, A., Cauchy, A., Giannakopoulos, C., Spano, D. & Briche, E. (2021) Investigating the climate-related risk of forest fires for Mediterranean Islands' blue economy. *Sustainability*, **13**, 10004.
Bahçekapılı, Y. & Yalçıntan, M.C. (2024) Emotions and Islandness: Exploring Interactions of Urban Activist Communities in the Prince Islands of Istanbul. *Crisis, Conflict and Celebration: Ethnographic Studies of European Cities* (ed. by K. Kajdanek, A. Bednarczyk and R. Carvalho), pp. 251-273. Springer Nature Singapore, Singapore.
Cadima Ribeiro, J.A., Vareiro, L., Remoaldo, P. & Monjardino, I.C. (2025) Residents' perceptions of the impacts of tourism in the Azores archipelago (Portugal): A cluster analysis. *Tourism and Hospitality Research*, **25**, 274-288.
Cavallaro, F. & Ciraolo, L. (2005) A multicriteria approach to evaluate wind energy plants on an Italian island. *Energy Policy*, **33**, 235-244.
Connell, J. (2018) Islands: balancing development and sustainability? *Environmental Conservation*, **45**, 111-124.
Cuka, A. (2025) Chapter 14 - Challenges and opportunities in implementing sustainable development on islands. *Integrated Planning for Sustainable Resilient Regions* (ed. by R.A. Castanho), pp. 219-234. Elsevier.
De Montis, A., Ledda, A., Serra, V., Manunta, A. & Calia, G. (2025) Urban Green Spaces and Climate Changes: Assessing Ecosystem Services for the Municipality of Sassari (Italy). *Land*, **14**, 1308.
Doménech-Sánchez, A., Laso, E. & Berrocal, C.I. (2021) Water loss in swimming pool filter backwashing processes in the Balearic Islands (Spain). *Water Policy*, **23**, 1314-1328.
Feliciano, D., Bouriaud, L., Brahic, E., Deuffic, P., Dobsinska, Z., Jarsky, V., Lawrence, A., Nybakk, E., Quiroga, S., Suarez, C. & Ficko, A. (2017) Understanding private forest owners' conceptualisation of forest management: Evidence from a survey in seven European countries. *Journal of Rural Studies*, **54**, 162-176.
Hof, A. & Schmitt, T. (2011) Urban and tourist land use patterns and water consumption: Evidence from Mallorca, Balearic Islands. *Land Use Policy*, **28**, 792-804.
Kallis, G., Stephanides, P., Bailey, E., Devine-Wright, P., Chalvatzis, K. & Bailey, I. (2021) The challenges of engaging island communities: Lessons on renewable energy from a review of 17 case studies. *Energy Research & Social Science*, **81**, 102257.

Kefalas, G., Kalogirou, S., Poirazidis, K. & Lorilla, R.S. (2019) Landscape transition in Mediterranean islands: The case of Ionian islands, Greece 1985–2015. *Landscape and Urban Planning*, **191**, 103641.

Leka, A., Lagarias, A., Stratigea, A. & Prekas, P. (2025) A Methodological Framework for Assessing Overtourism in Insular Territories—Case Study of Santorini Island, Greece. *ISPRS International Journal of Geo-Information*, **14**, 106.

Loftsdóttir, K. & Mixa, M.W. (2025) From Iceland to the Canary Islands: Understanding the Appeal of Mass Tourism in the Age of Over-Tourism. *Tourism and Hospitality*, **6**, 76.

Marathianou, M., Kosmas, C., Gerontidis, S. & Detsis, V. (2000) Land-use evolution and degradation in Lesvos (Greece): a historical approach. *Land Degradation & Development*, **11**, 63-73.

Mauger, R., Diestelmeier, L. & Nieuwenhout, C. (2024) Harnessing EU legal concepts for the energy transition on islands. *The Journal of World Energy Law & Business*, **17**, 167-183.

McElroy, J.L. (2003) Tourism development in small islands across the world. *Geografiska Annaler: Series B, Human Geography*, **85**, 231-242.

Meco, J.n., Guillou, H., Carracedo, J.-C., Lomoschitz, A., Ramos, A.-J.G. & Rodríguez-Yánez, J.-J. (2002) The maximum warmings of the Pleistocene world climate recorded in the Canary Islands. *Palaeogeography, palaeoclimatology, palaeoecology*, **185**, 197-210.

Morales, J., Rodríguez, A., Alberto, V., Machado, C. & Criado, C. (2009) The impact of human activities on the natural environment of the Canary Islands (Spain) during the pre-Hispanic stage (3rd–2nd Century BC to 15th Century AD): an overview. *Environmental Archaeology*, **14**, 27-36.

Moustakas, A., Zemah-Shamir, S., Tase, M., Zotos, S., Demirel, N., Zoumides, C., Christoforidi, I., Dindaroglu, T., Albayrak, T., Ayhan, C.K., Fois, M., Manolaki, P., Sandor, A.D., Sieber, I., Stamatiadou, V., Tzirkalli, E., Vogiatzakis, I.N., Zemah-Shamir, Z. & Zittis, G. (2025) Climate land use and other drivers' impacts on island ecosystem services: A global review. *Science of The Total Environment*, **973**, 179147.

Nurse, L.A., McLean, R.F., Agard, J., Briguglio, L.P., Duvat-Magnan, V., Pelesikoti, N., Tompkins, E. & Webb, A. (2014) Small islands. *Climate change 2014: Impacts, adaptation, and vulnerability. Part B: Regional aspects. Contribution of working group II to the fifth assessment report of the intergovernmental panel on climate change*, pp. 1613-1654.

Oikonomou, E.K., Kilias, V., Goumas, A., Rigopoulos, A., Karakatsani, E., Damasiotis, M., Papastefanakis, D. & Marini, N. (2009) Renewable energy sources (RES) projects and their barriers on a regional scale: The case study of wind parks in the Dodecanese islands, Greece. *Energy Policy*, **37**, 4874-4883.

Otto, R., Krüsi, B. & Kienast, F. (2007) Degradation of an arid coastal landscape in relation to land use changes in Southern Tenerife (Canary Islands). *Journal of Arid Environments*, **70**, 527-539.

Pathirana, A. (2025) Small islands: living laboratories revealing global climate and sustainable development challenges. *Frontiers in Climate*, **Volume 6 - 2024**

Pla-Rabes, S., Matias, M.G., Gonçalves, V., Vázquez Loureiro, D., Marques, H., Bao, R., Buchaca, T., Hernández, A., Giralt, S. & Sáez, A. (2024) Global warming triggers abrupt regime shifts in island lake ecosystems in the Azores Archipelago. *Communications Earth & Environment*, **5**, 571.

Rebolo-Ifrán, N., Lois, N.A. & Lambertucci, S.A. (2025) Wind energy development in Latin America and the Caribbean: Risk assessment for flying vertebrates. *Environmental Impact Assessment Review*, **112**, 107798.

Ricart, S., Villar-Navascués, R., Reyes, M., Rico-Amorós, A.M., Hernández-Hernández, M., Toth, E., Bragalli, C., Neri, M. & Amelung, B. (2024) Water–tourism nexus research in the Mediterranean in the past two decades: A systematic literature review. *International Journal of Water Resources Development*, **40**, 57-83.

Vicente-Serrano, S.M., Tramblay, Y., Reig, F., González-Hidalgo, J.C., Beguería, S., Brunetti, M., Kalin, K.C., Patalen, L., Kržič, A., Lionello, P., Lima, M.M., Trigo, R.M., El-Kenawy, A.M., Eddenjal, A., Türkes, M., Koutroulis, A., Manara, V., Maugeri, M., Badi, W., Mathbout, S., Bertalanič, R., Bocheva, L., Dabanli, I., Dumitrescu, A., Dubuisson, B., Sahabi-Abed, S., Abdulla, F., Fayad, A., Hodzic, S., Ivanov, M., Radevski, I., Peña-Angulo, D., Lorenzo-Lacruz, J., Domínguez-Castro, F., Gimeno-Sotelo, L., García-Herrera, R., Franquesa, M., Halifa-Marín, A., Adell-Michavila, M., Noguera, I., Barriopedro, D., Garrido-Perez, J.M., Azorin-Molina, C., Andres-Martin, M., Gimeno, L., Nieto, R., Llasat, M.C., Markonis, Y., Selmi, R., Ben Rached, S., Radovanović, S., Soubeyroux, J.-M., Ribes, A., Saidi, M.E., Bataineh, S., El Khalki, E.M., Robaa, S., Boucetta, A., Alsafadi, K., Mamassis, N., Mohammed, S., Fernández-Duque, B., Cheval, S., Moutia, S., Stevkov, A., Stevkova, S., Luna, M.Y. & Potopová, V. (2025) High temporal variability not trend dominates Mediterranean precipitation. *Nature*, **639**, 658-666.

Vogiatzakis, I.N., Balzan, M.V., Drakou, E.G., Katsanevakis, S., Padoa-Schioppa, E., Tzirkalli, E., Zotos, S., Álvarez, X., Külvik, M., Fonseca, C., Moustakas, A., Martínez-López, J., Mackelworth, P., Mandzukovski, D., Ricci, L., Srdjevic, B., Tase, M., Terkenli, T.S., Zemah-Shamir, S., Zittis, G. & Manolaki, P. (2023) Enhancing Small-Medium IsLands resilience by securing the sustainability of Ecosystem Services: the SMILES Cost Action. *Research Ideas and Outcomes*, **9**, e116061.

Zittis, G., Zoumides, C., Zemah-Shamir, S., Tase, M., Zotos, S., Demirel, N., Christoforidi, I., Dindaroğlu, T., Albayrak, T., Ayhan, C.K., Fois, M., Manolaki, P., Sandor, A., Sieber, I.M., Stamatiadou, V., Tzirkalli, E., Vogiatzakis, I.N., Zemah-Shamir, Z. & Moustakas, A. (2025) Insular ecosystem services in peril: a systematic review on the impacts of climate change and other drivers. *Climatic Change*, **178**, 127.